\documentclass[structabstract]{aa}  
\usepackage{amsmath}
\usepackage{graphicx}
\usepackage[varg]{txfonts}
\usepackage{longtable}
\usepackage{lscape}
\usepackage{txfonts}
\begin{document}
   \title{Optical spectroscopy of EX Lupi during quiescence and outburst} 

   \subtitle{Infall, wind, and dynamics in the accretion flow}

   \author{Aurora Sicilia-Aguilar\inst{1}, \'{A}gnes K\'{o}sp\'{a}l\inst{2,3}\thanks{ESA Fellow}, 
   Johny Setiawan\inst{4},   P\'{e}ter \'{A}brah\'{a}m\inst{5},
   Cornelis Dullemond\inst{4,6}, Carlos Eiroa\inst{1}, Miwa Goto\inst{4}, 
   Thomas Henning\inst{4}, \and Attila Juh\'{a}sz\inst{2}}

   \institute{\inst{1} Departamento de F\'{\i}sica Te\'{o}rica, Facultad de Ciencias, Universidad Aut\'{o}noma de Madrid, 28049 Cantoblanco, Madrid, Spain\\
              \email{aurora.sicilia@uam.es}\\
	      \inst{2} Leiden Observatory, Leiden University, PO Box 9513, 2300RA Leiden, The Netherlands\\
	      \inst{3} Research and Scientific Support Department, European Space Agency, ESA/ESTEC, SRE-SA, PO Box 299, 2200 AG, Noordwijk, The Netherlands\\
	      \inst{4} Max-Planck-Institut f\"{u}r Astronomie, K\"{o}nigstuhl 17, 69117 Heidelberg, Germany\\
 	      \inst{5} Konkoly Observatory of the Hungarian Academy of Sciences, PO Box 67, 1525 Budapest, Hungary\\
              \inst{6} Institut f\"{u}r Theoretische Astrophysik, Zentrum f\"{u}r Astronomie, Universit\"{a}t Heidelberg, Albert-Ueberle-Str. 2, 69120 Heidelberg, Germany
           }

   \date{Submitted December 1 2011,  accepted June 7 2012}

% \abstract{}{}{}{}{} 
% 5 {} token are mandatory
  \abstract
   {EX Lupi is the prototype of EXor variables. After 50 years of
   mild variability and smaller outbursts, the object again experienced a 
   large outburst in 2008. }
   {We explore the accretion mechanisms in EX Lupi during its
   pre-outburst, outburst, and post-outburst phases.}
   {We analyze 10 high-resolution optical spectra of EX Lupi, taken
   before, during, and after the 2008 outburst.
   In both quiescence and
outburst, the star presents many permitted emission lines. These include lines 
typical of accreting T Tauri stars, 
plus a large number of neutral and ionized
metallic lines (mostly Fe~I and Fe~II).  During the outburst, the number of emission
lines increases to about a thousand, and their structure
shows a narrow and a broad component (NC and BC).  We study the
structure of the BC, which is highly variable on short timescales (24-72h).}
   {An active chromosphere can explain the metallic
   lines in quiescence and the outburst NC. The dynamics of the BC line profiles suggest
that these profiles originate in a hot, dense, non-axisymmetric, and
non-uniform accretion 
column that suffers velocity variations along the line-of-sight 
on timescales of days. 
Assuming Keplerian rotation, the emitting region 
would be located at $\sim$0.1-0.2 AU, which is consistent with the location of the inner disk rim,
but the velocity profiles of the lines reveal a combination of
rotation and infall. Line ratios of ions and neutrals
can be reproduced assuming a temperature of T$\sim$6500~K for electron densities 
of a few times 10$^{12}$cm$^{-3}$ in the line-emitting region.
The line profiles also indicate that there is an accretion-related inner disk wind.}
{The data confirm that the 2008 outburst was an episode of increased 
accretion, albeit much stronger than previous EX Lupi and 
typical EXors outbursts. The line profiles are consistent
with the infall/rotation of a non-axisymmetric structure that
could be produced by clumpy accretion during the outburst phase. A strong inner disk
wind appears in the epochs of higher accretion.
The rapid recovery of the system after the
outburst and the similarity between the pre-outburst and post-outburst
states suggest that the accretion channels are similar during the whole
period, and only the accretion rate varies, providing a superb environment for studying the accretion
processes.  }


\authorrunning{Sicilia-Aguilar et al.}

\titlerunning{EX Lupi optical spectroscopy}

   \maketitle
%
%________________________________________________________________

\section{Introduction}
EX Lupi is a young, low-mass star (spectral type M0;
Herbig 1977; Gras-Vel\'{a}zquez \& Ray 2005), prototype of
the EXor family among young variable T Tauri stars (Herbig 1989). Its
irregular variability has been known for about 120 years
(McLaughlin 1946), although the causes of these variations
and their association with young stars were not suggested
until much later (Herbig 1950). While the normal variability
pattern is characterized by irregular fluctuations, with small
outbursts where the brightness increases by $\sim$2.5 mag
over timescales of months to few years, a much 
stronger outburst was observed in 1955-57 (Herbig 1977). This
episode was followed
by more than 50 years of small-scale variability, until in January 2008
another large outburst was observed (Jones 2008).
Spectroscopic observations of EX Lupi in outburst and during its
quiescence phases revealed that it is a M0 classical
T Tauri star (CTTS) that suffers episodes of variable mass
accretion (Herbig et al. 2001). The larger outbursts
could be assimilated to episodes of largely increased mass infall,
up to three orders of magnitude over the typical quiescence levels
(Juh\'{a}sz et al. 2012; K\'{o}sp\'{a}l et al. 2011).

The 2008 outburst offered a remarkable opportunity to check the hypothesis
of increased accretion episodes, and to study the
causes of mass infall and its effect on the disk. 
The outburst was characterized by a rapid rise from the quiescence
state in November 2007, to a maximum around January 2008 (Jones 2008).
It was followed by a slow decay until it reached quiescence levels
again by September 2008. During this decrease, the optical magnitude
of the object suffered
a few oscillations with amplitudes 1-2 mag (increasing with time) and 
period of about 35 days (Juh\'{a}sz et al. 2012).
During the 2008 outburst, the optical spectrum
of EX Lupi became full of broad metallic emission lines, with hardly any photospheric
absorption line remaining visible (K\'{o}sp\'{a}l et al. 2008;
Aspin et al. 2010). This was an outstanding behavior, given that the emission 
lines observed in previous, weaker outbursts were rather narrow
and more similar to the emission lines observed during
quiescence (Herbig et al. 2001; Lehmann et al. 1995).
The brightening of the star also affected the disk, 
allowing a real-time experiment on silicate processing and
crystallization as well as mixing within the disk (\'{A}brah\'{a}m
et al. 2009; Juh\'{a}sz et al. 2012) that suggested the development
of a disk wind during the outburst phase. CO observations
uncovered the rotational pattern of the emitting gas (Goto et al. 2011),
which is consistent with gaseous material filling in the inner hole 
of the dust disk that had been
suggested by observations of the object in quiescence (Sipos et al.
2009). These observations were consistent with the results of additional near infrared
(near-IR) spectroscopy data, which revealed the presence of a $\sim$0.1-0.3 AU dust-free
region in the disk, where gas may be located at the edge of the disk
and/or in the outer part of a funnel flow (K\'{o}sp\'{a}l et al. 2011).

Here we present and analyze a collection of ten optical spectra of EX Lupi taken
during the pre-outburst, 2008 outburst, and post-outburst
phases. Optical spectroscopy traces the innermost structure
in the system, being the key to understanding the
underlying physics of the EXor outburst processes. It 
allows us to determine the differences in accretion,
chromospheric emission, and innermost gaseous disk structure between
the quiescence and the outburst phase.  
In Section \ref{obs}, we describe the observations and data reduction.
In Section \ref{analysis}, we study the emission features observed
in quiescence (before and after the outburst) and during the outburst. In Section \ref{discussion},
we explore the physical processes and structures that
can be related to the optical emission lines, including
accretion, wind, and chromospheric activity. We finally summarize
our results in Section \ref{conclu}.

\section{Observations and data reduction\label{obs}}

%\begin{landscape}
\begin{table*}
\caption{Journal of observations and estimated magnitudes} 
\label{obs-table}
\begin{tabular}{l c c c c c l}
\hline\hline
 Date & JD & Instrument & Exp. Time (s) & S/N   & Magnitude & Comments \\
\hline
2007 July 28 	& 2454309.114	&  FEROS & 1600	& 0.7/30/28	& 12.7:$^1$ & Pre-outburst	\\
2007 July 29 	& 2454310.156 	&  FEROS &  1800	& 0.3/30/20	& 12.7:$^1$ & Pre-outburst	\\
2007 July 30 	& 2454311.187	&  FEROS &  1800	& 0.2/15/14	& 12.7:$^1$ & Pre-outburst	\\
2008 April 21 	& 2454577.277   &  FEROS &  2400	& 13/220/65	& 9.55$^2$  & Outburst \\
2008 May 5 	& 2454591.426	&  FEROS &  2000	& 1/46/23	& 9.7:$^1$  & Outburst	\\
2008 May 6 	& 2454592.152	&   FEROS & 2000	& 5/90/36	& 9.7:$^1$  & Outburst	\\
2008 May 8 	& 2454594.360	&   HARPS & 2000	& ---/32/---	& 9.7$^1$   & Outburst	\\
2008 June 16 	& 2454633.084	&   FEROS & 2000	& 8/165/51	& 9.6$^1$   & Outburst	\\
2010 June 9 	& 2455356.339	&   FEROS & 2000	& 0.1/13/8	& 13.7$^1$  & Post-outburst	\\
2010 June 11 	& 2455358.271	&   FEROS & 2000	& 0.2/10/8	& 13.7:$^1$ & Post-outburst	\\
\hline
\end{tabular}
\tablefoot{Information on the spectroscopy observations. The signal-to-noise ratio (S/N) is measured 
in three different parts of the spectra ($\sim$3650, $\sim$6700, and $\sim$8800\AA). 
The magnitudes are visual data from AAVSO (marked with $^1$) and
V magnitude taken with WFI at the 2.2m telescope in La Silla (marked with $^2$; Juh\'{a}sz et al.
2012). If no data was obtained on the given day, we interpolated from nearby observations and
indicate interpolated observations with ``:". Variations in the S/N between spectra taken at similar
brightness of the source are due to weather conditions. The pre-outburst data were presented in
Sipos et al. (2009).}
\end{table*}
%\end{landscape}

A total of 10 spectra were taken with the Fiber-fed Extended Range Optical Spectrograph (FEROS), 
mounted on the 2.2m Max-Planck Gesellschaft/European Southern
Observatory (MPG/ESO) telescope, and the High Accuracy Radial velocity Planet Searcher (HARPS), 
mounted on the 3.6m telescope
(both located at ESO-La Silla observatory)
during MPG guaranteed time. FEROS has a resolution of $R=48\,000$
and a wavelength coverage from 3700--9200 \AA\ (Kaufer et al. 2000). HARPS 
has a similar resolution, but a shorter wavelength coverage (3800--6900 \AA).
The observations took place during 2007, 2008, and 2010. Three of them correspond
to the pre-outburst phase (2007 July 28, 29, and 30), 5 were taken close to
the maximum of the 2008 outburst and in the subsequent months
(on 2008 April 21, May 5, 6, and 8, June 16),
and two more were obtained in 2010, again during a quiescent phase of
EX Lupi (2010 June 9, 11). 
A journal of the observations is presented in Table \ref{obs-table}.
The AAVSO reports visual magnitudes of 12.7 for the pre-outburst observations
(extrapolating from measurements in the previous week), 10.1 for 2008 April 21 
(with our WFI observations giving V=9.554$\pm$0.003, see Juh\'{a}sz et al. 2011),
9.7 and 9.6 for 2008 May 8 and 2008 June 16, with similar values extrapolating
to 2008 May 5 and 6, and a magnitude of 13.7 on 2010 June 9. Therefore, the
star was slightly brighter than its minimum during our pre-outburst observations, 
and reached its typical minimum magnitude at the time the post-outburst 
spectra were taken.
The spectral reduction was performed using the online data reduction
pipeline installed at the telescope. This pipeline produced a one-dimensional
spectrum from 370--920 nm, generated by merging 39 echelle-order spectra.
The line analysis (including measurements of wavelength, equivalent width [EW], 
and integrated intensity)
was done with standard tasks within the IRAF\footnote{IRAF is distributed by the National Optical Astronomy Observatories,
which are operated by the Association of Universities for Research
in Astronomy, Inc., under cooperative agreement with the National
Science Foundation.} \textit{onedspec} package.

\section{Analysis \label{analysis}}

%\longtab{2}{
\onllongtab{2}{
\begin{landscape}
\begin{longtable}{lccccccl}
\caption{\label{quiline-table} Lines observed during the quiescence phase (pre- and post-outburst).
 The spectra of the pre- and post-outburst phases, respectively, are combined to 
increase the S/N in the measured values, given the lack of significant variability. 
 For each line and observations, we give the observed wavelength 
($\lambda$) EW (\AA) and FWHM (km/s). Typical accuracy is 0.02\AA\
in wavelength, and 0.1 \AA\ in EW.} \\
\hline\hline
 Line $\lambda_{lab}$ & E$_k$  & A$_{ki}$ &2007 $\lambda_{obs}$  &2007 EW/FWHM  &2010 $\lambda_{obs}$ &2010 EW/FWHM  & Comments\\
(\AA) & (eV) &  (s$^{-1}$) & (\AA) & (\AA / km/s)& (\AA) & (\AA / km/s) & \\ 
\hline
\endfirsthead
\caption{Continued.}\\
\hline\hline
 Line $\lambda_{lab}$ & E$_k$  & A$_{ki}$ &2007 $\lambda_{obs}$  &2007 EW/FWHM  &2010 $\lambda_{obs}$ &2010 EW/FWHM  & Comments\\
(\AA) & (eV) &  (s$^{-1}$) & (\AA) & (\AA / km/s)& (\AA) & (\AA / km/s) & \\ 
\hline
\endhead
\hline
\endfoot
H10 3797.97 & 13.46 & 7.12e+04 & 3797.94 & -6.6 / 182 & --- & --- & \\   
H9 3835.35 & 13.43 & 1.22e+05 & 3835.33 & -2.5 / 98 & --- & --- & \\   
H8 3889.06 & 13.39 & 2.21e+05 & 3888.91 & -8.5 / 150 & 3888.97 & -2.6 / 238 & \\ 
H$\epsilon$ 3970.04 & 13.32 & 4.39e+05 & 3969.96 & -5.6 / 120 & 3970.01 & -2.5 / 72 & \\ 
H$\delta$ 4101.71 & 13.22 & 9.73e+05 & 4101.72 & -11.9 / 142 & 4101.66 & -6.1 / 119 & \\ 
H$\gamma$ 4340.47 & 13.05 & 2.53e+06 & 4340.45 & -11.0 / 133 & 4340.43 & -9.6 / 124 & \\ 
H$\beta$ 4861.28 & 12.75 & 8.42e+06 & 4861.25 & -17.1 / 148 & 4861.26 & -14.4 / 130 & \\ 
H$\alpha$ 6562.57 & 12.09 & 5.39e+07 & 6562.53 & -32.2 / 187 & 6562.6 & -21.5 / 151 & \\
\hline
\end{longtable}
\end{landscape}
}

%\longtab{3}{
\onllongtab{3}{
\begin{longtable}{lcccl}
\caption{\label{outline-table} Lines observed during the outburst phase, 
grouped according to element.\\
Lines classified as `INDEF' could not be
found within the available line databases. The comments include `W' (weak line), `S' (strong line),
`Wer' (line gets weaker with time), `Ser' (line gets stronger with time),
`NC' (only the NC is clearly detected), `BC' (only the BC is detected),
`M' (the line appears as a shoulder, merged with another stronger line), 
`B' (the line is a blend of several lines that cannot be disentangled),
 `S/N' (low S/N), `P' (the line has a clean profile, not being
surrounded by other stronger lines), `Wi' (the profile shows blueshifted absorption consistent with
an origin in a wind-like component), `Wi-I' (the line profile displays evidence of wind that turned into
infall at later stages of the outburst),`Br' (remarkably broad line), `T' (line is contaminated by telluric absorption),
`OA' (only observed in April), `WA' (weak in April), `SA' (strong in April), `WJ' (became weak in June), 
`SJ' (became stronger in June),`NJ' (not observed in June), `MJ' (line merged to a stronger one in June),
`OJ' (only observed in June), `OQ' (only observed in quiescence),
`p' (most probable classification, in the case of lines with two potential identifications, e.g.
pFeI means that the line is probably the Fe~I one, from the two listed), 
Lines marked with : have an uncertain classification.
Lines named INDEF could not be identified within the NIST database. In some cases, there is a nearby
line (listed in the comments), but the offset in wavelength is much larger than typically observed.
Excitation potentials and transition probabilities not listed in the NIST database are left blank.
Some lines not listed in the NIST database but classified by Appenzeller et al.(1986) are identified
with the comment AJJ86.  } \\
\hline\hline
 Line $\lambda_{lab}$ &E$_k$  & A$_{ki}$  & $\lambda_{obs}$  & Comments \\ 
 (\AA) & (eV) & (s$^{-1}$) & (\AA) & \\
\hline
\endfirsthead
\caption{Continued.}\\
\hline\hline
 Line $\lambda_{lab}$ &E$_k$  & A$_{ki}$  & $\lambda_{obs}$  & Comments \\ 
 (\AA) & (eV) & (s$^{-1}$) & (\AA) & \\
\hline
\endhead
\hline
H~I 3659.41 & 13.59 & 1.77e+02 & 3659.41 & BC \\     
H~I 3679.37 & 13.57 & 1.70e+03 & 3679.38 & BC \\     
H~I 3691.551 & 13.56 & 3.69e+03 & 3691.53 & BC \\     
H~I 3697.157 & 13.55 & 4.91e+03 & 3697.1 & BC \\     
H~I 3711.98  & 13.54 & 9.2e+03 & 3712.21 & BC \\     
H~I 3750.151 & 13.5 & 2.83e+04 & 3750.24 & BC \\    
H$\delta$ 4101.71 & 13.22 & 9.73e+05 & 4101.0 & BC \\
H$\gamma$ 4340.47 & 13.05 & 2.53e+06 & 4339.8 & BC \\
\hline
\end{longtable}
}

Young stars such as CTTS and Herbig Ae/Be (HAeBe) stars are characterized
by the presence of numerous emission lines in their spectra 
(see for instance Appenzeller et al. 1986; Hamann \& Persson 1992a,b; 
Hamann 1994). Many of the emission lines observed in EX Lupi are
common among young T Tauri stars (TTS) and HAeBe stars,
although the 
large number of metallic emission lines found
in EX Lupi (both in quiescence and outburst) are
rare among typical CTTS, having precedents only in the observations
of some very active CTTS and other EXors (Appenzeller et al. 1986; 
Hamann \& Persson 1992a; Beristain et al. 1998; Parsamian \& Mujica 2004; Herbig 2008).
The commonly observed permitted lines include those of H~I emission (especially the Balmer
series dominated by H$\alpha$), the Ca~II IR triplet, and He~I emission, all of 
which are typically related to the accretion activity. In some
cases, some O I lines and the Na I doublet appear as well in emission.
Forbidden lines (e.g. [O~I], [S~II]) are also a common occurrence in CTTS and HAeBe
(Hamann 1994), although EX Lupi does not display any evidence of them during
either quiescence or outburst, confirming the results of K\'{o}sp\'{a}l
et al. (2011) in the near-IR. This suggests that,
despite the wind signatures we describe later, there is no shock
or it is quenched (Nisini et al. 2005). 
Therefore, the environment around EX Lupi seems to be clean of cloud
material and remnant protostellar envelopes, in agreement with model
predictions (Sipos et al. 2009).

The observed emission lines are listed in
Table \ref{quiline-table} (for the quiescence
phase) and Table \ref{outline-table} (for
the outburst phase). 
The lines were identified using
the National Institute of Standards and Technology (NIST) 
database\footnote{http://physics.nist.gov/PhysRefData/ASD/lines$\_$form.html} 
(Ralchenko et al. 2010), 
which includes information about the wavelength, 
excitation potential (or energy of the upper level, E$_k$), 
multiplicity, relative intensity, and transition 
probabilities (or Einstein coefficients, A$_{ki}$). A few lines could not be identified in the NIST database,
being thus listed as ``INDEF" in our Tables. Some of the
lines that do not appear in the NIST database 
are listed in Appenzeller et al. (1986), in which cases we adopted
their classification. In the following sections, we analyze
the observed lines during the quiescence and outburst phases.

\subsection{The quiescence phases: Pre- and post-outburst \label{pre}}

During the quiescence phases, EX Lupi displayed the photospheric absorption
lines characteristic of a late-type star, together with several broad
and narrow emission lines (see Sipos et al. 2009
and Table \ref{quiline-table}).
There were no significant differences between the metallic lines 
observed in the pre-outburst and post-outburst quiescence phases,
although the post-outburst spectra have poorer signal-to-noise ratio (S/N), 
which resulted in fewer lines
being clearly identified. 
Only the accretion-related lines (H~I Balmer series, He~I,
Ca~II) display significant epoch and day-to-day variations.

\begin{figure}
   \centering
   \resizebox{\hsize}{!}{\includegraphics[width=\textwidth]{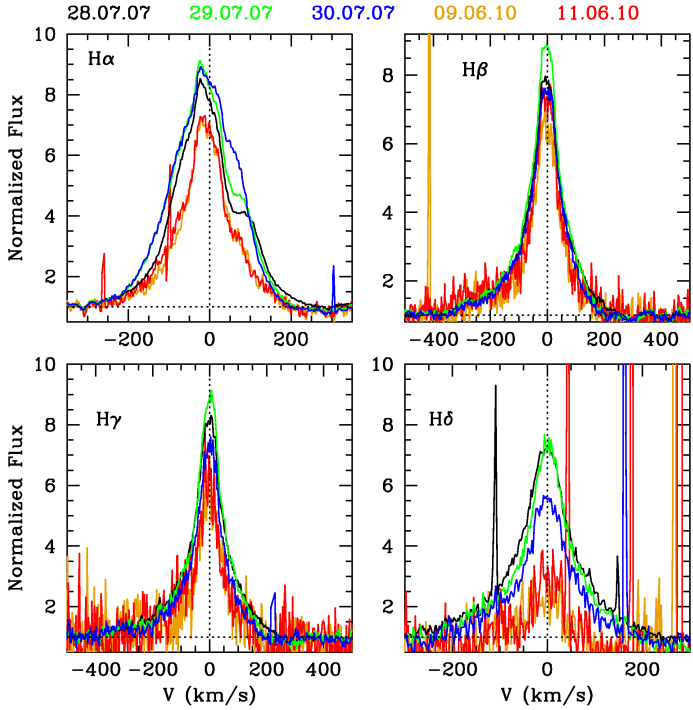}}
\caption{H lines during the quiescence phases (pre- and post-outburst).
The zero velocity is marked by a dotted line. Different colors represent
the observations on different dates (black for 2007 July 28, green for 2007 July 30,
blue for 2007 July 30, orange for 2010 June 9, and red for 2010 June 11).
\label{Hqui-fig}}
\end{figure}

\begin{figure}
   \centering
   \resizebox{\hsize}{!}{\includegraphics[width=\textwidth]{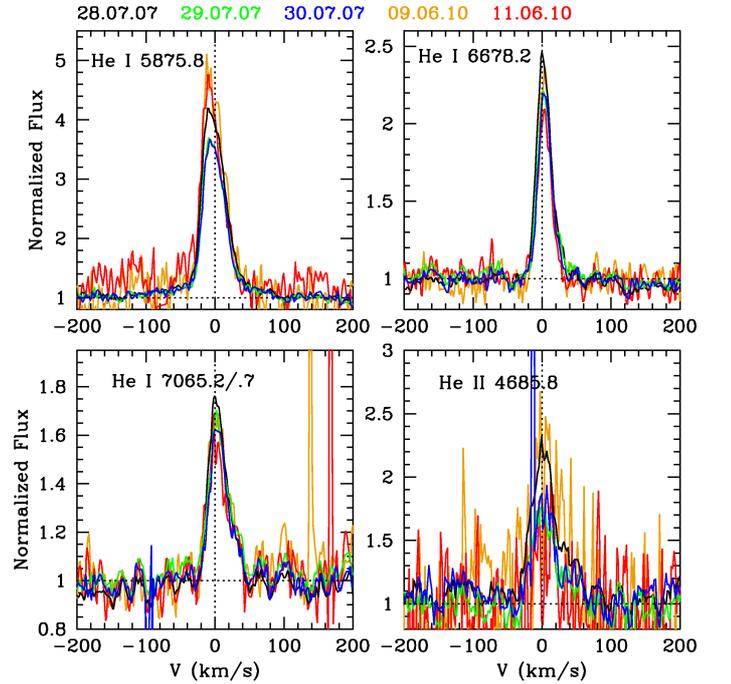}}
\caption{He lines during the quiescence phases (pre- and post-outburst).
The zero velocity is marked by a dotted line. Different colors mark different
dates as in Figure \ref{Hqui-fig}. \label{Hequi-fig}}
\end{figure}

The H~I Balmer series in emission is the most remarkable feature, 
with H$\alpha$ being the strongest line in the spectra (Figure \ref{Hqui-fig}).
The H$\alpha$ line appears to be asymmetric, with redshifted absorption that was most
prominent during the pre-outburst phase and caused
the peak of the line to appear blueshifted by about 20 km/s. Redshifted
absorptions signatures of infall that are commonly seen in CTTS, and 
had also been observed during the 1993 EX Lupi outburst (Lehmann et al.
1995), when H$\alpha$ displayed an inverse P-Cygni profile. The day-to-day variations
in the H$\alpha$ emission line suggest that there was a mild variation in the accretion rate. The line wings extend
beyond $\pm$200 km/s, with the red part being in general broader. The higher Balmer
lines are more symmetric and less variable, with the redshifted absorption being present 
to a lesser extent. They have wings extending beyond
$\pm$200 km/s. A
small blueshifted absorption centered at $\sim$-250 km/s is seen in H$\beta$, 
which is an indication of a weak cool wind. 

The quiescence spectra also display strong He~I emission at 5878, 6678, and 7065 \AA\
 (Figure \ref{Hequi-fig}).
The lines are narrow, with a full width at half maximum (FWHM) of $\sim$50 km/s 
and asymmetric, with a stronger red component,
and mild variability in both profile and intensity.
Given the high temperatures and densities required for the formation of He~I lines, 
they probably originate in the densest and hottest parts of the accretion column,
close to the star, and in the accretion shock itself (Hamann \& Person 1992a; Berinstain et al. 2001). 
The He~II 4686\AA\ line is also detected, as 
noted by Herbig (2007) during other epochs. He~II lines require extremely hot
conditions for their production, so that they are rare even in the proximity
of O stars, and more typical of novae, Wolf-Rayet stars, and planetary nebulae 
(Torres \& Massey 1987; Mason et al. 2010).

\begin{figure}
   \centering
   \resizebox{\hsize}{!}{\includegraphics{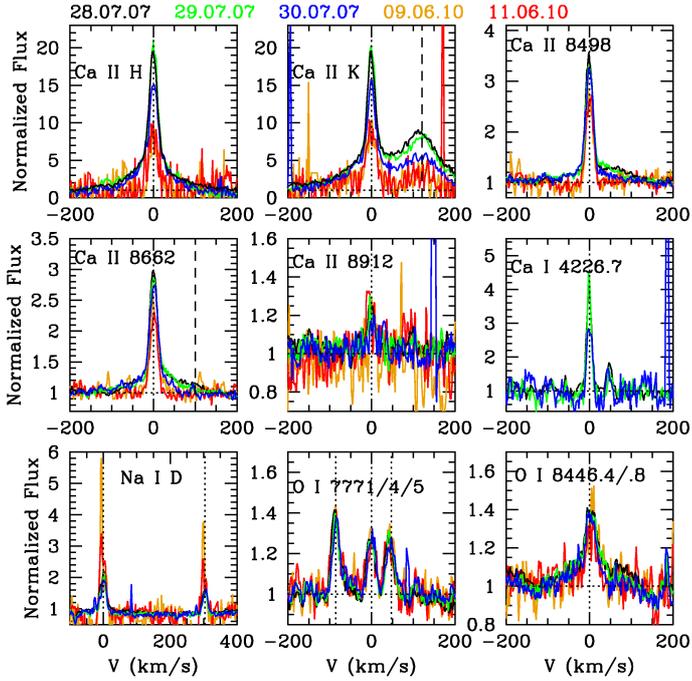}}
\caption{Some of the Ca~I, Ca~II, Na~I, and O~I lines seen in emission during the quiescence
phases (pre- and post-outburst). The zero velocity is marked by a dotted line
(in the cases of the Na~I~D and O~I triplet, we mark the zero velocity for each resolved line).
The dashed lines in the Ca~II K and Ca~II 8662\AA\ lines mark the position of the Balmer H$\epsilon$ and
Paschen 13 lines, respectively. The weaker line redwards of Ca~I 4226 is Fe~I 4227.
Different colors mark different dates as in Figure \ref{Hqui-fig}. \label{Otherqui-fig}}
\end{figure}

The Ca~II H and K lines, together with the Ca~II IR triplet\footnote{The
Ca~II line at 8542\AA\ falls at the edge of the FEROS gap, so is not visible in
quiescence and only partially visible during the outburst phase. We therefore
excluded it from the figures.}, are also observed
in emission (see Figure \ref{Otherqui-fig}). The Ca~II lines have wings
extending to $\pm$100 km/s and display a weak asymmetric profile consistent with infall.
The Na I D is also observed in emission, with a weak
blueshifted absorption, and the O~I triplet at 7771/4/5\AA\ and O~I line at 8446 \AA\
also appear in emission.

\begin{figure}
   \centering
   \resizebox{\hsize}{!}{\includegraphics{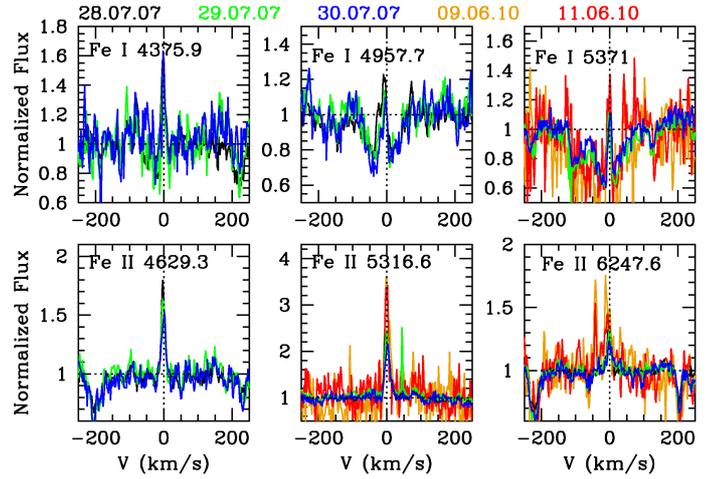}}
\caption{Some of the Fe~I and Fe~II lines seen in emission during the quiescence phases (pre- and
post-outburst). The zero velocity is marked by a dotted line. 
Different colors mark different dates as in Figure \ref{Hqui-fig}. \label{Fequi-fig}}
\end{figure}

Several Fe~I and Fe~II lines are observed in emission (Figure \ref{Fequi-fig}), 
as noted in previous weaker outbursts (Lehmann et al. 1995; Herbig et al. 2001).
Herbig (2007) reported the presence of Fe~I, Fe~II, Ti~II, and
other narrow emission lines during the quiescence period around 2000. 
We confirm all these lines in our data and, in addition, 
we observe some Ca~I, Cr~I, Cr~II, Mg~I, Si~I, 
Si~II, Mn~I, and Co~I emission lines.
The neutral metallic lines appear frequently superimposed on the corresponding 
photospheric absorption lines, resembling in some cases double-peaked absorption
lines due to the weak and narrow emission component in the center.
The spectral resolution does not allow us to study in great detail the 
line profiles, which appear symmetric and narrow (FWHM$\sim$10-20 km/s). 
The emission lines observed in the post-outburst phase
tend to have lower E$_k$ than those observed in the pre-outburst, which could
be indicative of a mild change in the temperature of the emitting region
towards lower temperatures (maybe due to a lower accretion rate), although the poorer S/N in
the post-outburst may increase the difficulty of the identification of the weakest lines.

\subsection{The outburst \label{outburst}}

During the outburst phase, the spectrum became a forest of strong emission lines
(K\'{o}sp\'{a}l et al. 2008) with large daily variations
in the lines and continua.
The lines correspond to the same species detected
in quiescence, but the number of neutral and ionized 
permitted metallic lines increased from $\sim$150 in quiescence, to
several hundreds and probably over a thousand in outburst\footnote{We estimate that 
a similarly large number of lines to those indicated in
Table \ref{outline-table} could not be identified
owing to blends with nearby features.} with typical intensities 
that were 10-100 times higher.
All the emission lines, except H~I, contain 
a broad and a narrow component (BC and NC).
Composite NC+BC line profiles have been 
reported in previous outbursts (Herbig 2007), as well as during the earliest 
phases of the 2008 outburst (Aspin et al. 2010), although most CTTS
show either broad or narrow components (Hamann \& Persson 1992a). The lines reported
by Herbig (2007) for the 1998 outburst had inverse P-Cygni profiles,
which are typical of infall. The lines observed here show rapidly variable and complex 
signatures, both blueshifted and redshifted, as we later discuss.
The BCs typically appear to be blended due to the number of broad lines. Therefore, the line
identification during the outburst phase was always done based on the
NCs, with widths $\sim$10-20km/s.  The NCs
always appear at rest velocity ($\sim$0$\pm$3 km/s),
do not suffer strong variations in their profile (although they may vary in strength), 
and are thus similar, albeit stronger, to the lines observed in quiescence.
Table \ref{profi-table}
lists the line parameters for the handful of lines that do not suffer
strong blending.

\begin{figure}
   \centering
   \includegraphics[width=5cm]{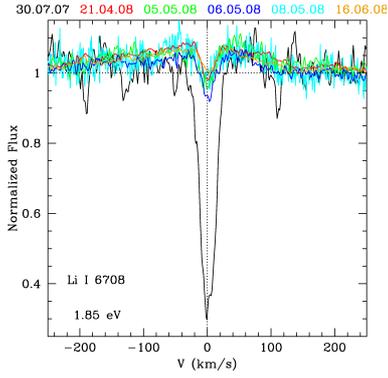}
\caption{Li I 6708\AA\ line, the only one that still appears in
absorption during the outburst phase. Colors mark the observations at different
dates (red for 2008 April 21, green for 2008 May 5, blue for 2008 May 6, cyan for 2008 May 8,
orange for 2008 June 16, and black for the scaled spectrum in quiescence).\label{LiIout-fig}}
\end{figure}

\addtocounter{table}{1}

The very veiled Li I 6708\AA\ line
(Figure \ref{LiIout-fig}) is the only one 
seen in absorption in the outburst spectra,
although it also appears to contain some weak BC emission. 
All photospheric lines typical of low-mass
stars (Fe~I, Ca~I, Mg~I, K~I) are seen in emission. In addition,
we see the accretion-related lines typical of CTTS (H~I Balmer and Paschen 
series, He~I, Ca~II IR triplet) together with ionized lines that are common in
cases of strong winds and novae (Fe~II, Cr~II, Ti~II; Mason et al. 2010) and 
C~I lines (Hamann \& Persson 1992a,b). 

\begin{figure}
   \centering
   \resizebox{\hsize}{!}{\includegraphics{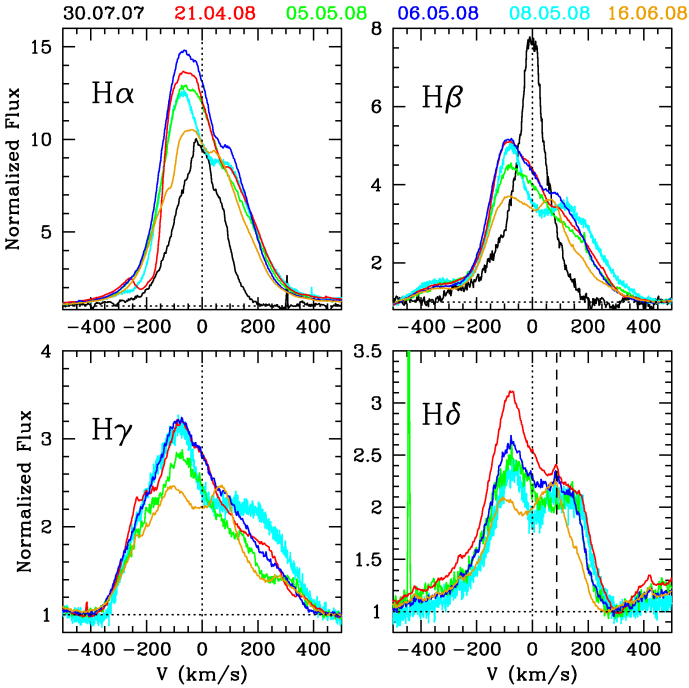}}
\caption{Some H lines during the outburst phase. The zero velocity is marked by a dotted line.
For comparison, the scaled quiescence spectrum of 2007 July 30 is displayed (except for H$\gamma$
and H$\delta$, owing to low S/N).  
The color code is as in Figure \ref{LiIout-fig}.\label{Hout-fig}}
\end{figure}

\begin{figure}
   \centering
   \resizebox{\hsize}{!}{\includegraphics{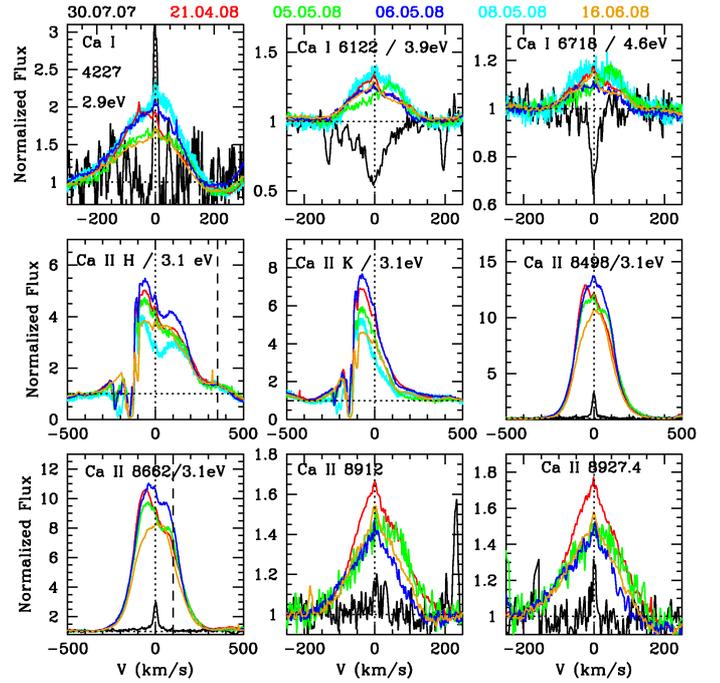}}
\caption{Some of the Ca~I and Ca~II lines observed in emission during the outburst phase.
The zero velocity is marked by a dotted line. In the Ca~II H panel,
we also indicate with a dashed line the Fe~II 3938.3\AA\ line (4.81 eV),
and note that the redshifted emission may be contaminated by more
Fe~I and Fe~II lines in this wavelength range.
For comparison, the quiescence spectrum of 2007 July 30 is displayed.
The color code is as in Figure \ref{LiIout-fig}.
\label{Caout-fig}}
\end{figure}

The H$\alpha$ emission line is the strongest and broadest one in the spectra, 
followed by the rest
of the H~I Balmer series (Figure \ref{Hout-fig}) and other typical 
accretion-related features (Ca~II H and K lines, Ca~II IR triplet;
see Figure \ref{Caout-fig}). The broad line wings extend
beyond $\pm$400 km/s (for the H~I Balmer series) and $\pm$200 km/s (in the 
other accretion-related lines). No NC is visible in the H lines.
H$\alpha$ and H$\beta$ display a redshifted absorption feature
centered on $\sim$+20-50 km/s,
while the structures of the H$\gamma$ and H$\delta$ lines are more complex, and vary from
redshifted absorption in most observations, to nearly double-peaked lines in June. 
H$\alpha$ and H$\beta$ also display strong blueshifted
absorption at -200 $-$ -250 km/s that can be identified with an
accretion-related wind. The blueshifted absorption varies with time, decreasing in parallel with
the accretion rate from April to June, which is consistent with the
general picture of mass loss related to the accretion processes in young stars
(e.g. Shu et al. 1994; Calvet et al. 1997). The observations of Aspin et al.(2010) also
confirm that there was a decrease in the strength of the blueshifted absorption as the outburst
faded.

\begin{figure}
   \centering
   \resizebox{\hsize}{!}{\includegraphics{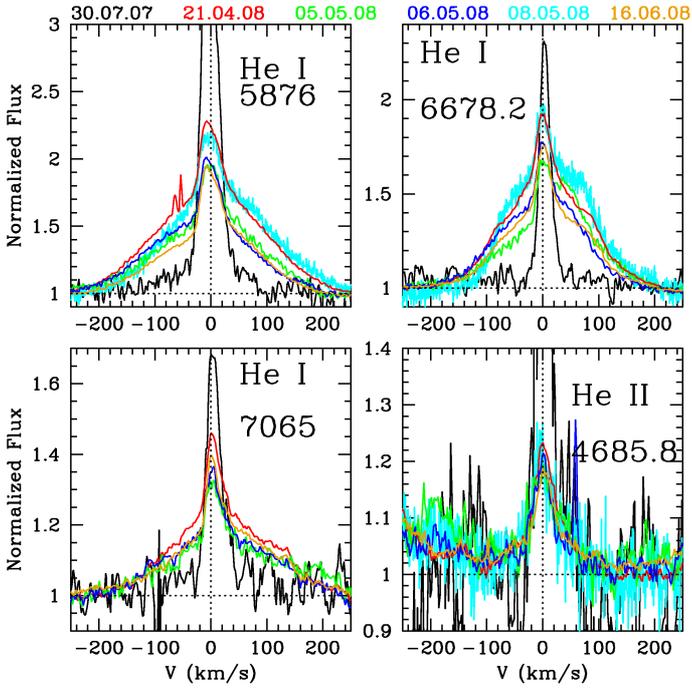}}
\caption{Some of the He lines observed during the outburst phase. The zero velocity is marked by a dotted line.
For comparison, the quiescence spectrum of 2007 July 30 is displayed.
The color code is as in Figure \ref{LiIout-fig}.\label{Heout-fig}}
\end{figure}

Several He~I lines are also observed (Figure \ref{Heout-fig}). 
The high temperatures and densities required to excite the 
He~I transitions suggest that they 
are produced at the bottom of the accretion column. Nevertheless, we 
observe a BC+NC structure in the He~I lines, which indicates that these 
lines may have two
different origins (e.g. chromosphere and accretion column). The BC of the He~I
lines are asymmetric, with flattened wings that give them a triangular appearance, 
which differs from the profiles observed in H I and in the metallic lines. The NC in the He~I lines is
broader ($\sim$40 km/s) than the typical NC observed in the metallic lines, which has
been suggested as proof that the lines originate in a transition zone between
the chromosphere and corona, rather than have a chromospheric origin 
(Hamann \& Persson 1992a). 
He~II emission at 4686\AA\ is also observed. 
The He~II line is relatively narrow and
similar to the He~I NC, with which it probably shares a physical location.

\begin{figure}
   \centering
   \resizebox{\hsize}{!}{\includegraphics{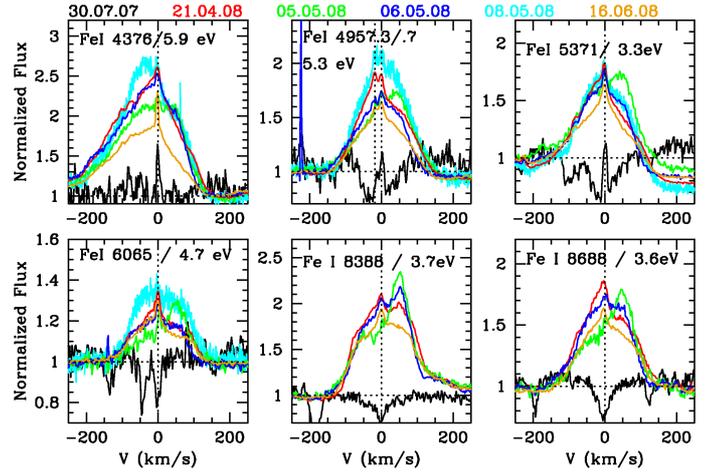}}
\caption{Some of the Fe~I lines observed in emission during the outburst phase. 
The zero velocity is marked by a dotted line.  
For comparison, the quiescence spectrum of 2007 July 30 is displayed.
The color code is as in Figure \ref{LiIout-fig}.\label{FeIout-fig}}
\end{figure}

The most abundant neutral metallic lines observed in emission in the
outburst are, as in quiescence, 
those of Fe~I (see Figure \ref{FeIout-fig}). 
There are no strong Fe~I lines with E$_k >$7 eV, with most
of the lines having E$_k\sim$3.5-6.8 eV,
in agreement with the results of K\'{o}sp\'{a}l et al. (2011) for
the metallic lines in the near-IR spectrum.
The BC of the Fe~I lines are very asymmetric. 
Close to the outburst maximum in April, the lines tend to
have two absorption components, one that is blueshifted ($\sim$-50 km/s) and 
another that is redshifted (broader, resulting
in the steep profile in the red part of the line), 
which could be the signature of wind and infall, respectively
(see for instance Fe I 4376 in Figure \ref{FeIout-fig}).
The line profiles are similar, although have a stronger NC, a stronger redshifted 
absorption, and weaker or absent blueshifted absorption, in June 16 (with
profiles reminiscent of high-inclination magnetospheric
infall; Hartmann et al. 1994).
During the May 2008 observations, we observe strong dynamical changes 
in the Fe~I line profiles that are not correlated with the accretion rate
and have timescales of 1-2 days. The BCs appear
strongly redshifted on 2008 May 5, and the redshifted emission subsequently
decreases (2008 May 6) until both the redshifted and the blueshifted part of the line
are roughly equal or even blue-dominated (by 2008 May 8). This is consistent
with the double-peaked lines reported by Aspin et al. (2010), although 
our more complete coverage suggests that the double peaks are a dynamical effect
rather than due to the particular geometry of a static structure.
Remarkably, although there is some similarity to the Fe lines observed
by Beristain et al. (1998) in DR Tau, which also display a NC+BC structure, the 
dynamics observed here are in strong contrast to the stable velocities observed
in DR Tau.

\begin{figure}
   \centering
   \resizebox{\hsize}{!}{\includegraphics{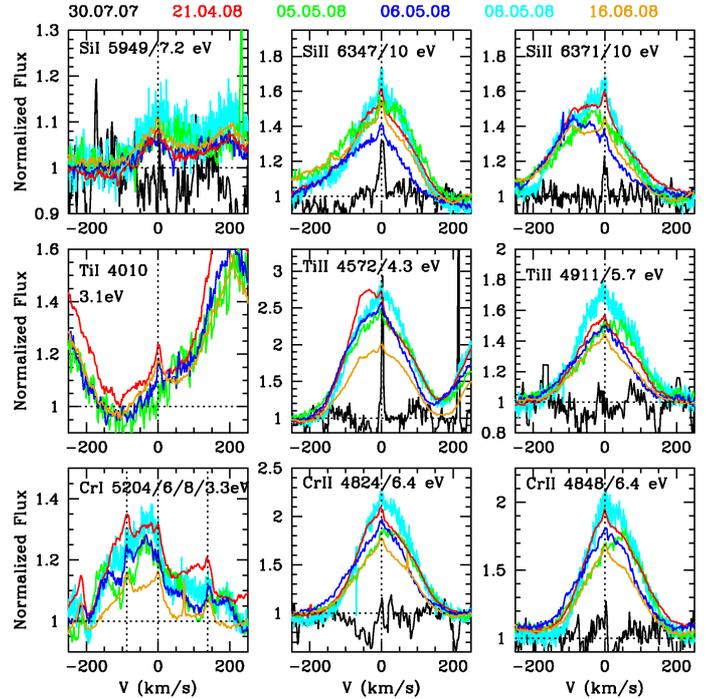}}
\caption{Some of the Si~I, Si~II, Ti I, Ti~II, Cr I, and Cr II lines observed in emission 
during the outburst phase. 
The zero velocity is marked by a dotted line. For comparison, the quiescence spectrum of 2007 July 30 is displayed
at the wavelengths where there is enough S/N.
\label{SiTiCrout-fig}}
\end{figure}

\begin{figure}
   \centering
   \resizebox{\hsize}{!}{\includegraphics{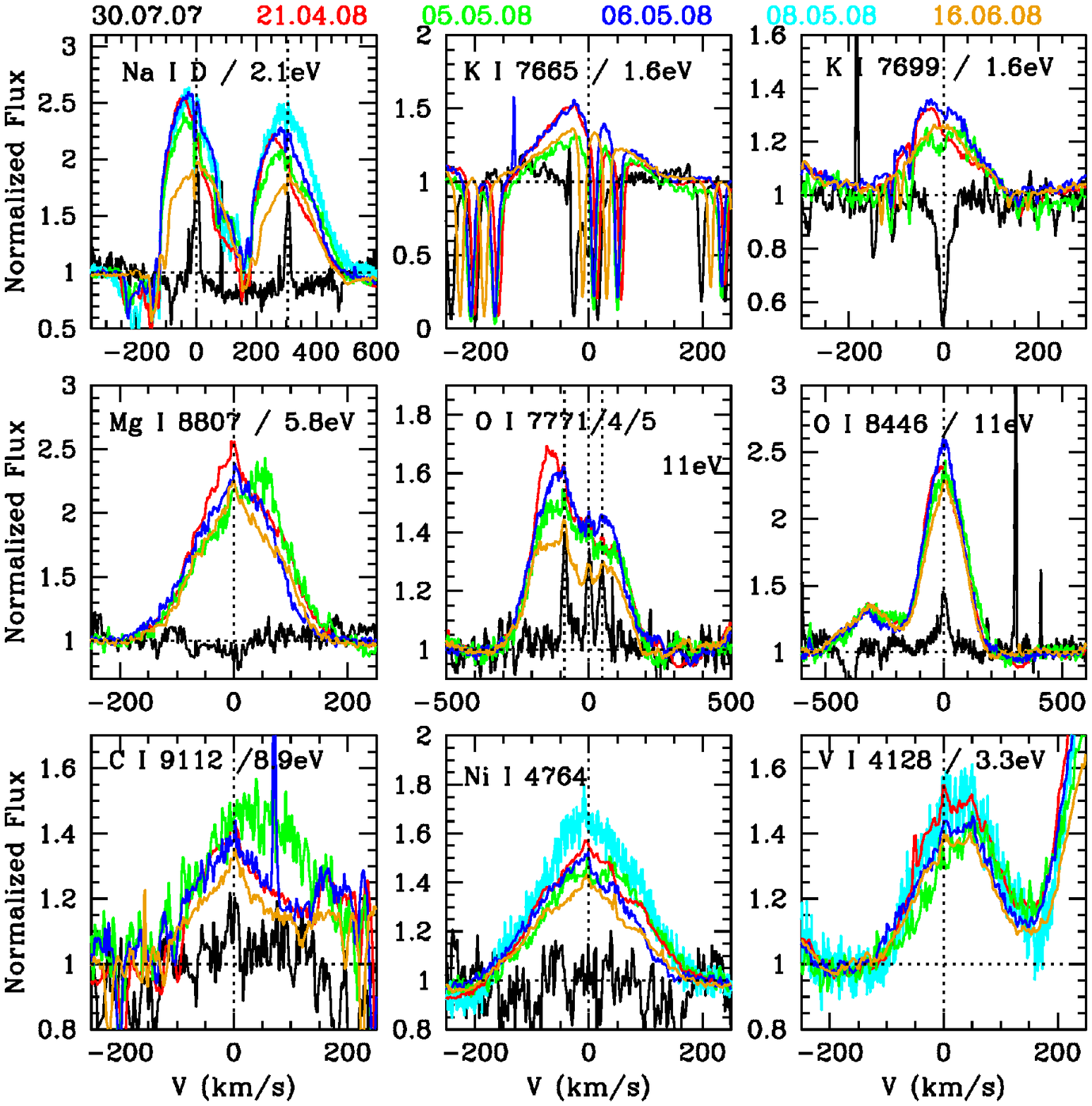}}
\caption{Examples of other lines observed in emission during the outburst phase. 
The zero velocity is marked by a dotted line.
For comparison, the quiescence spectrum of 2007 July 30 is displayed.
The color code is as in Figure \ref{LiIout-fig}.\label{Otherout-fig}}
\end{figure}

Other neutral lines observed include Ca~I, 
Co~I, Si~I, Cr~I, 
Ni~I, Mg~I, Ti~I, and V~I (see Figures \ref{Caout-fig}, \ref{SiTiCrout-fig}, \ref{Otherout-fig}). 
The NC of the
metallic lines strengthens in June as the accretion rate drops and the BC become
weaker, which is probably a contrast effect.
A Na I line at 7113 \AA\, (E$_k$=4.9 eV) 
also seems to be present, in addition to the accretion-related
Na I D. There is no clear detection of Mn I during the outburst phase,
probably owing to blends with stronger lines.
Towards the red part of the spectrum, some K I and C I lines are detected in emission,
although they are strongly affected by telluric absorption (Figure \ref{Otherout-fig}). 
All of these lines show the dynamical variations observed in the
Fe~I lines, with similar velocity shifts, except for the Ca~I 4227\AA\ line,
whose profile remains very stable.
This line has a low E$_k$ (2.9 eV), which could suggest an origin in a distinct cooler region,
but other lines with E$_k\sim$3-3.5 eV display the typical dynamical pattern described above. 
The maximum observed E$_k$
varies for each species, being E$_k\leq$5 eV for Ca~I, E$_k\leq$3 eV for 
Ti~I, E$_k\leq$4.2 eV for Co~I, 
E$_k\leq$7.5 eV for Mg~I, E$_k\leq$4.2 eV for K~I, E$_k\leq$7.5 eV for Si~I, 
E$_k\leq$6 eV for Ni~I, E$_k\leq$3.7 eV for 
Cr~I, and E$_k\sim$8.9-9.1eV for the C~I lines.

\begin{figure}
   \centering
   \resizebox{\hsize}{!}{\includegraphics{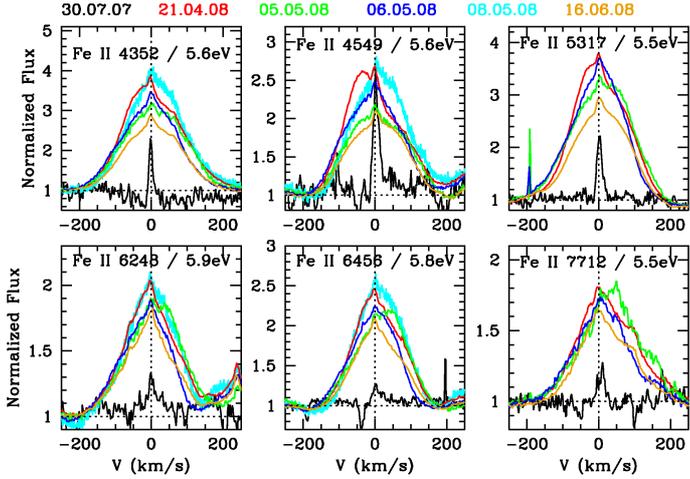}}
\caption{Some of the Fe~II lines seen in emission during the outburst phase. 
The zero velocity is marked by a dotted line.
For comparison, the quiescence spectrum of 2007 July 30 is displayed.
The color code is as in Figure \ref{LiIout-fig}.\label{FeIIout-fig}}
\end{figure}

Fe~II and Ti~II are the most common ionized lines found in the spectra. They are usually very strong and 
follow the dynamical patterns observed in neutral lines, although their profiles are
affected by stronger absorption features. The April 2008
observations, which are the closest to the outburst
maximum, provide evidence of both blueshifted and redshifted absorption,
which are signatures of
wind and infall, respectively, while the data from June 16 2008 is more consistent with infall 
alone (Figures \ref{FeIIout-fig} and \ref{SiTiCrout-fig}). 
This is indicative of a combined origin for the lines in the accretion column and the accretion-related wind.
The E$_k$ of the ionized lines have typical energies 
of the upper level $\sim$5.5 eV for Fe~II and $\sim$4 eV
for Ti~II. We also observe some Si~II and Cr~II lines (with E$_{k}$$\sim$10 and $\sim$6.5 eV;
Figure \ref{SiTiCrout-fig}).
There is no evidence of any other ionized lines.
The relatively higher excitation potentials suggest formation in a
region with higher temperatures and/or lower electron densities than the places
where the neutral lines arise, as we discuss later on.

Although the Ca~II IR
triplet and the Ca~II H and K lines follow the accretion/wind pattern observed in the
H~I Balmer lines,
they also contain a NC centered at zero velocity, as seen in the
rest of the metallic lines (figure \ref{Caout-fig}). In contrast, other Ca~II lines
(e.g. Ca~II 8912 and 8927\AA) display the same 
dynamical imprint as the rest of metallic lines.
Finally, we also observe some O~I lines (7771/4/5\AA\ triplet and 8446\AA\ line;
Figure \ref{Otherout-fig}). The O~I triplet 
appears to be strongly blueshifted, as has been seen in
stars with strong accretion and wind emission (Hamann \& Persson 1992b).

\section{Discussion  \label{discussion}}

\subsection{Accretion and winds \label{accretionwind} }

The broad and asymmetric profiles of the emission lines observed in young CTTS (especially,
those of the H I Balmer series) are usually interpreted in terms of magnetospheric
accretion and winds (K\"{o}nigl 1991; Muzerolle et al. 1998, 2001;
Appenzeller et al. 1986; Hamann \& Persson 1992b; Alencar \& Basri 2000).
Other mechanisms capable of producing the observed features in CTTS, such as extended winds and jets alone 
and emission from low-temperature H I gas (Kurosawa et al. 2006; Whelan et al. 2004;
Bary et al. 2008) can be ruled out in the case of EX Lupi by noting the
redshifted absorption and the ratio of Br$\gamma$ to Pa$\beta$ observed in the infrared (IR),
consistent with  high temperature H I
(Sipos et al. 2009; K\'{o}spal et al. 2011).
The accretion-related lines observed in the 2008 EX Lupi
outburst have wings extending up to relatively large velocities
($\pm$400 km/s) compared to other CTTS and to simple magnetospheric accretion
models. Higher temperatures in the accretion flow contribute to increasing the widths of
H$\alpha$, especially when combined with high accretion rates (Kurosawa et al. 2006). 
The inclusion of wind components combined with detailed infall
models can produce profiles in good agreement with our observations (e.g. Lima et al. 2010).  
While H$\alpha$ and H$\beta$ can be
interpreted in terms of accretion, wind, and infall, the profiles of the
higher Balmer lines (which are more optically thin, and thus trace deeper/denser structures) 
reveal multi-component variable
profiles. The stronger redshifted absorption present in the higher Balmer lines 
is indicative of an origin closer
to the star. The weakening of the blueshifted absorption in the higher Balmer lines is
expected if they originate in an optically thin wind (Hartmann et al. 1994). 

It is remarkable that the velocities
observed for the blueshifted and redshifted absorption components in the H~I Balmer lines
do not significantly differ in outburst and quiescence (although the
blueshifted absorption component in quiescence is very weak). This suggests that the
mechanisms and paths for both infall and outflow are not significantly different between outburst
and quiescence, and only the amount of accreted and expelled matter varies. 
The post-outburst phase appears very similar to the pre-outburst,
which implies that the system made a rapid recovery after the outburst, which is also consistent
with the idea of stable accretion/outflow paths and mechanisms.
 
To make a simple estimate of the accretion rates during the pre-outburst, outburst,
and post-outburst phases, we measured the 10\% velocity wings  of the H$\alpha$
line (H$\alpha_{10\%}$), which were on average 345,
550, and 300 km/s, respectively. Following Natta et al.
(2004), the H$\alpha_{10\%}$ correlates to the accretion rate (in M$_\odot$/yr) 
according to

\begin{eqnarray}
\log(dM/dt) = -12.9(\pm0.3) + 9.7\times 10^{-3} (\pm0.7) H\alpha_{10\%}.
\label{eq:1}
\end{eqnarray}

We thus derive average
accretion rates of 3$\times 10^{-10}$M$_\odot$/yr,
3$\times 10^{-8}$M$_\odot$/yr, and 1$\times 10^{-10}$M$_\odot$/yr during the pre-outburst,
outburst, and post-outburst phase. 
The EW of the Li I 6708\AA\ line, the only one clearly present in absorption
during the outburst, agrees with these results, being 
0.46, 0.04, and 0.6 during the pre-outburst, outburst, and post-outburst 
phases, respectively. These values and their variations are
consistent with accretion-related veiling. 
During the outburst phase, the H$\alpha_{10\%}$
decreases from $>$570 km/s
to $\sim$540 km/s between April and June, which is
consistent with a variation in the rate from $>$4$\times 10^{-8}$M$_\odot$/yr
to $\sim$2$\times 10^{-8}$M$_\odot$/yr.
Although the spread in this relation is large,
mostly due to line-of-sight effects,
the observations suggest an increase in the accretion rate
of about two orders of magnitude between quiescence and outburst, and a factor of a few
higher accretion rate in the pre-outburst phase than in the post-outburst epoch.
The accretion rates derived from Brackett $\gamma$ by Sipos et al. (2009)
and Juh\'{a}sz et al. (2012) reveal an increase in \.{M} up to 3 orders
of magnitude from quiescence to outburst
(from $\sim$4$\times 10^{-10}$ to 2$\times 10^{-7}$M$_\odot$/yr), 
which is consistent with our results, given that our H$\alpha$ measurement
in April is a lower limit\footnote{Note that the accretion estimate in April is affected
by the strong blueshifted absorption, so it can only be considered as
a lower limit. To minimize the effect of the wind, if we assumed the line to be relatively symmetric, the resulting
H$\alpha_{10\%}$ would be on the order of 630-650 km/s,
corresponding to an accretion rate of 2-3$\times 10^{-7}$M$_\odot$/yr.}.
The line profile variations of H$\alpha$ also agree with this picture
(Figure \ref{Hout-fig}).

The presence of a wind appears in many TTS to be correlated to the accretion activity,
with the mass loss due to wind being approximately 10\% of the accretion rate
(Calvet 1997). Although typical EXor outbursts are
characterized by line profiles dominated by infall (inverse P Cygni profiles; Herbig 2007),
the 2008 EX Lupi outburst presents strong evidence of a powerful wind,
as noted by Aspin et al. (2010). The January spectra
from Aspin et al. (2010) appear to detect a stronger wind absorption (below the
continuum), which suggests that there was a higher accretion and outflow at earlier stages of the
outburst, followed by a progressive decrease in the infall rate and wind absorption with time.
The main evidence of a wind component comes from both the blueshifted
absorption in H$\alpha$, H$\beta$, and the metallic lines, especially the ionized ones (Fe~II, Ti~II), 
observed in April
2008 and also reported in January 2008 by Aspin et al. (2010). 
The wind signatures are weaker in subsequently acquired spectra, being
much weaker (or absent) in our May and June 2008 observations. 
The blueshifted absorptions in the ionized metallic lines correspond to smaller velocities
than in the H~I lines (up to -100 km/s
instead of -200 km/s), which may be a combined effect of optical thickness
and element abundance in the wind.

The O~I line at 8446\AA\ is strong and relatively symmetric 
(Figure \ref{Otherout-fig}). On the other hand, 
the O~I 7771/4/5\AA\ triplet is consistent with
the presence of a powerful wind and infall. In April 2008, the peak of the BC of the
O~I 7771\AA\ line appeared both stronger than the NC and blueshifted
by about 60 km/s, and the line contained both blueshifted and 
redshifted absorption components. The blueshifted peak progressively
transformed into a shoulder as the accretion rate decreased,
probably owing to the wind becoming optically thinner. The infall-related redshifted
absorption also decreased as the accretion rate becomes lower. Further evidence of
blueshifted absorption is also found in the Na I D lines (Figure \ref{Otherout-fig}). 

This suggests that the 2008 outburst was extraordinary not only for EX Lupi, but
also for the class of the EXors. The presence of a
wind is a clear sign that the
accretion rate exceeded the typical rates measured in 
EXor outbursts (Herbig et al. 2001), although the
depth of the blueshifted absorption at its maximum is far from that typically 
seen in FUors (Hartmann \& Kenyon 1996).
The presence of a powerful wind
can also explain the transport of crystalline material to the outer disk,
as suggested by Juh\'{a}sz et al. (2012) based on the variations in the 
mineralogy of the outer disk. In this case, the wind must extend beyond the
relatively dense and warm regions observed by our optical spectra. 

\subsection{Dynamics of the metallic line-emission region: Stellar chromosphere and 
accretion column \label{metallic} }

%\begin{landscape}
\begin{table*}
\caption{Central line velocities  } 
\label{velocity-table}
\begin{tabular}{l c c c c c}
\hline\hline
 Elements & Apr. 21  & May 5 & May 6 & May 8 & June 16\\
\hline
Fe~I, Ca~I, Mg~I 		& 1.7$\pm$2.0(20) & 22.2$\pm$2.3(20) & -1.5$\pm$2.0(20) & 8.7$\pm$2.0(18) & 0.3$\pm$1.9(20) \\
Fe~II, Si~II, Ti~II, Cr~II 	& 5.0$\pm$2.3(29) & 16.4$\pm$3.2(29) & -4.5$\pm$2.9(29) & 11.5$\pm$2.5(26) & 3.5$\pm$2.7(29) \\
All & 3.3$\pm$1.4(60) & 16.8$\pm$2.0(60) & -3.5$\pm$1.6(60) & 10.2$\pm$1.6(48) & 2.7$\pm$1.5(60) \\
\end{tabular}
\tablefoot{Velocities derived from the Gaussian fit of the lines in Table\ref{profi-table}.
For each date and type of lines, we give the average velocity with standard deviation and the number of points
used in the calculation (in brackets). Only the lines that appear independent and are well-identified
(listed in Table\ref{profi-table}) have been taken into account in our calculations.}
\end{table*}
%\end{landscape}

\begin{figure*}
   \centering
   \includegraphics[width=18cm]{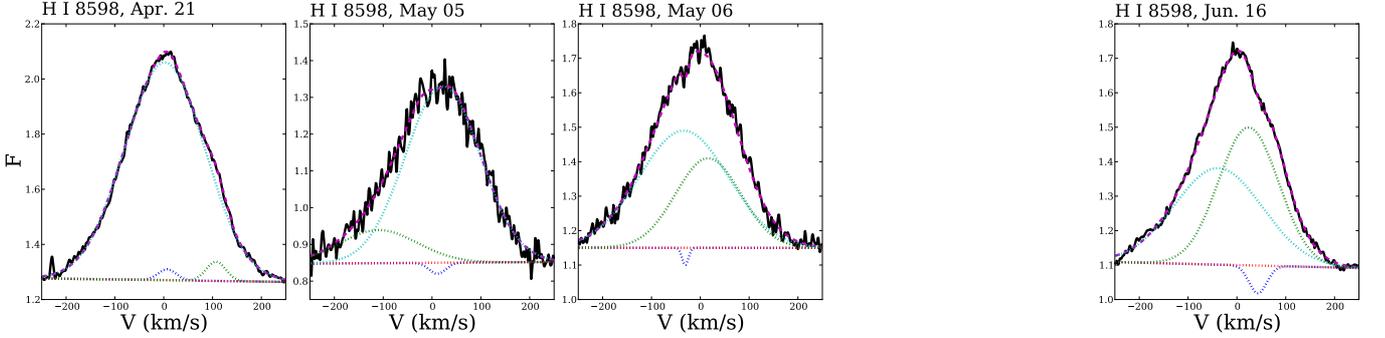}
\caption{H~I 8598\AA\ line: Results of the fit with three Gaussian components
for the different outburst epochs. The original data
is shown as the bold line, individual Gaussians
are marked by dotted lines, and the final fit is shown as a dashed line.   \label{Hfit-fig}}
\end{figure*}

\begin{figure*}
   \centering
   \includegraphics[width=18cm]{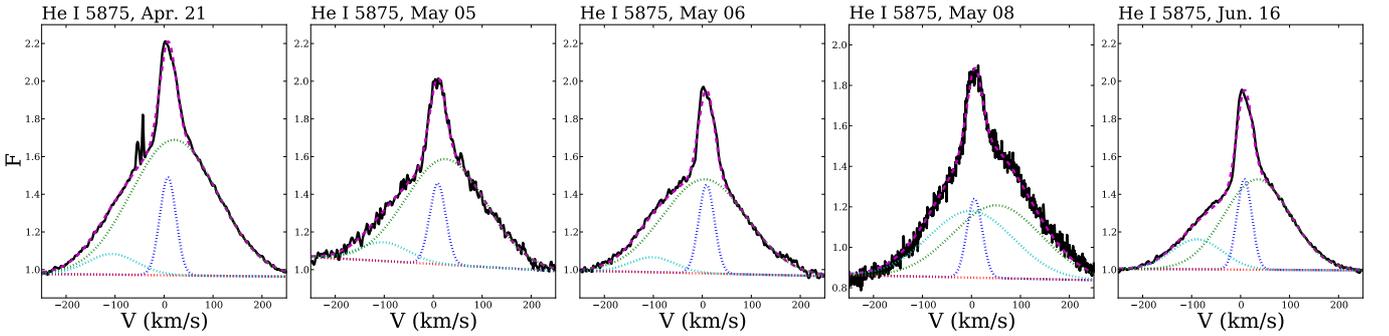}
\caption{He~I 5875\AA\ line: Results of the fit with three Gaussian components
for the different outburst epochs. The original data
is shown as the bold line, individual Gaussians
are marked by dotted lines, and the final fit is shown as a dashed line.   \label{Hefit-fig}}
\end{figure*}

As we have mentioned before, in the quiescence phase, all the metallic lines are narrow (FWHM$\sim$10-20 km/s), and 
frequently appear superimposed on the corresponding
(broader) photospheric absorption lines. The NCs of the metallic lines observed
in the outburst phase strongly resemble the quiescence emission lines.
In both cases, the lines appear centered on zero velocity and we can exclude
velocity variations of over 3 km/s in individual lines, which is consistent with
an origin very close to the stellar photosphere. 
The lines appear relatively symmetric and do not suffer significant
profile variations, although their intensities change.
The presence of weak metallic emission lines is relatively common among T Tauri stars
(Hamann \& Persson 1992a), and they have also been observed in other EXors
(Herbig 2008), although the case of EX Lupi in quiescence is remarkable 
because of the number and strength of the lines. The small velocities 
 and the relatively low excitation potentials agree with the lines being produced in an active chromosphere 
(Hamann \& Persson 1992a). This picture is consistent with the 
observations of high ultraviolet and soft X-ray variability: the
high X-ray absorption, which does not correspond to the observed optical
A$_V$, suggests a large column density in thick, dense accretion
flows in an object with a very active chromosphere (Grosso et al. 2010).

The situation during the outburst phase is
complex, given the strongly variable BC profiles in the metallic lines. 
Aspin et al. (2010) had already noticed variations in the line profile between
their January and May 2008 observations. Our more-complete dataset
reveals the interesting dynamics in the BC of the Fe~I, Ca~I, K I, Mg I, Ni I,
and V I lines,which is also confirmed by the weaker Si~I and Ti~I lines
present in the spectra.
The BC of the neutral metallic lines suffers a day-to-day shift, with peak 
variations of up to +50 km/s. The shift is also seen in
the ionized lines, although to a lesser extent, maybe due to the presence of
stronger absorption features.

Rapid variations have been observed in accretion-related
lines of very active TTS (e.g. RW Aur; Hartmann 1982; Alencar et al. 2005;
DR Tau; Alencar et al. 2001), which are typically associated
with non-steady accretion and outflow.
More rarely, these variations are also found in the metallic lines of very active CTTS
(e.g. SCrA; Appenzeller et al. 1986; Hamann \& Persson 1992a), although these
variations tend to affect the intensity rather than the velocity components
of the lines (e.g. Beristain et al. 1998). 
In Herbig Ae stars/UXors, velocity
variations in the metallic absorption lines have been identified
as infalling cold clumps of matter (Mora et al. 2002, 2004).
In general, rapid variations in the emission lines are most accurately explained by
global motions of large parcels of hot gas around the star.
The dynamical variations in the metallic lines differ
from the line profile variations owing to the variable accretion and wind observed in the
H~I Balmer series. This suggests that the H~I lines and the metallic lines
are produced (or dominated) by emission in physically distinct regions,
although excitation conditions and optical thickness may also play a role.
For instance, in regions with relatively low temperature ($<$7000 K), metallic
lines are expected to dominate over H I emission, while the H~I lines are produced
in regions with higher temperatures (Beristain et al. 1998).

\begin{figure*}
   \centering
   \includegraphics[width=18cm]{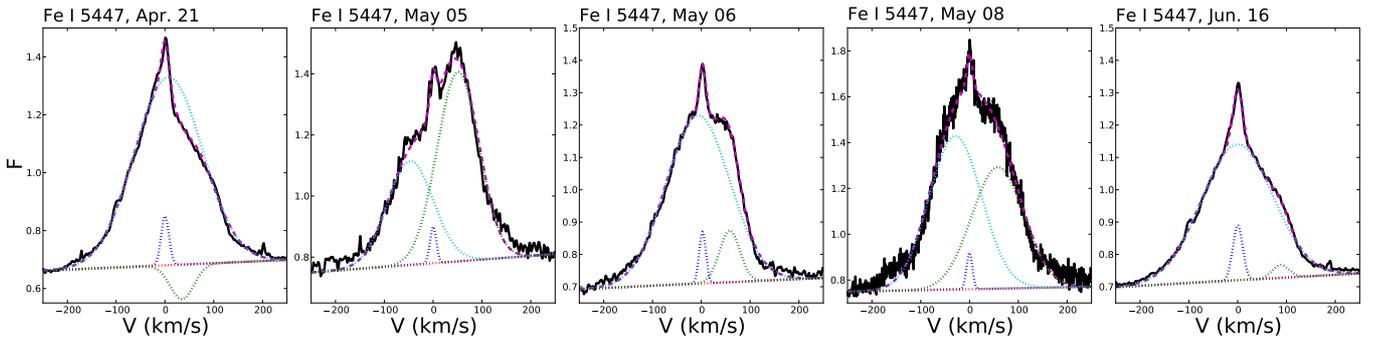}
\caption{Fe~I 5447\AA\ line: Results of the fit with three Gaussian components
for the different outburst epochs. The original data
is shown as the bold line, individual Gaussians
are marked by dotted lines, and the final fit is shown as a dashed line.   \label{FeI5447fit-fig}}
\end{figure*}

To determine the origin of the variability in the metallic lines, 
we analyzed in detail the magnitude of
the shift for the lines that do not appear to be blended 
(Table \ref{profi-table}). For each of
these lines, we measured the central wavelength and FWHM using IRAF task $splot$
by fitting a Gaussian profile to the
lines. We find that all the lines vary consistently from day to day. 
We observe a clear variation of the central line velocity
between approximately rest velocity and +22 km/s (Table \ref{velocity-table}).
The variations in the peak position of the BC are more extreme, and
span up to +50 km/s. 

To get a more detailed picture of the line structure, we performed
a line fit with the IRAF task $ngaussfit$ using three Gaussian components
for a selected sample of lines thought to have different origins (see the Appendix
for more details), including H I, He I,
and a large number of non-blended metallic lines. 
We avoided the Balmer series since their profiles are highly complex
and differ from all the other lines. 
Some of the best fits are displayed
in Figures \ref{Hfit-fig} to \ref{FeII5276fit-fig}, with more lines
being discussed in the Appendix. This exercise reveals at once clear differences
between H I, He I, and the metallic neutral/ionized lines. The most remarkable
features are the lack of NC in the H I lines, and the lack of dynamical signatures
in both the H I and He I lines, which trace the hotter and denser parts of the
flow/shock region. The profiles of H I and He I lines are very stable,
compared to the metallic lines. The dynamical signatures observed in the neutral and ionized
metallic lines are very similar and strongly correlated.

The three Gaussian fits confirm that the NC velocities are essentially
constant in time (within our level of precision), with the He I NC being the broadest, followed by the 
NC of the ionized lines
and finally, the NC of the neutral metallic lines. This is in good agreement with a 
relatively stable chromosphere where the lines with the higher excitation potentials
are produced in hotter, more turbulent regions. Alternative explanations for an origin of narrow
He I lines in the post-shock environment
have also been proposed (Beristain et al. 2001), although we do not measure
consistent redshifts in the He I lines. We also confirm that the
BC of H I and He I lines have very different profiles compared to the 
metallic lines, as expected if they have physically different origins within the
accretion column. In general, the H I and He I lines are more symmetric and,
although some redshifted and blueshifted components are observed, they never reach
the velocity shifts observed in the metallic lines, suggesting that the dynamical
changes observed in the latter occur in a cooler region.

The 2008 May 5, 6, and 8 observations of the neutral metallic lines can be very
well-reproduced with a narrow (FWHM$\sim$10-20\AA) Gaussian component for the NC, plus two broader
Gaussians with variable center velocity and width.  The  strongly redshifted component in the
2008 May 5 observations of the neutral and ionized 
metallic lines becomes very evident with the three Gaussian models, and we confirm
the velocity offset of $\sim$50 km/s. The velocity offset is
maximal on 2008 May 5, decreases by May 6, and the line is clearly more
symmetric by May 8. The corresponding observations of ionized metallic lines
are also interpretable in the same terms, but the velocity offsets vary. In particular,
although the maximum shift appears on May 5 for both neutral and metallic lines
and has the same velocity in both groups,
the ionized lines are much more symmetric on May 6 than the neutral lines,
which is also indicative of a slightly different physical origin.

\begin{figure*}
   \centering
   \includegraphics[width=18cm]{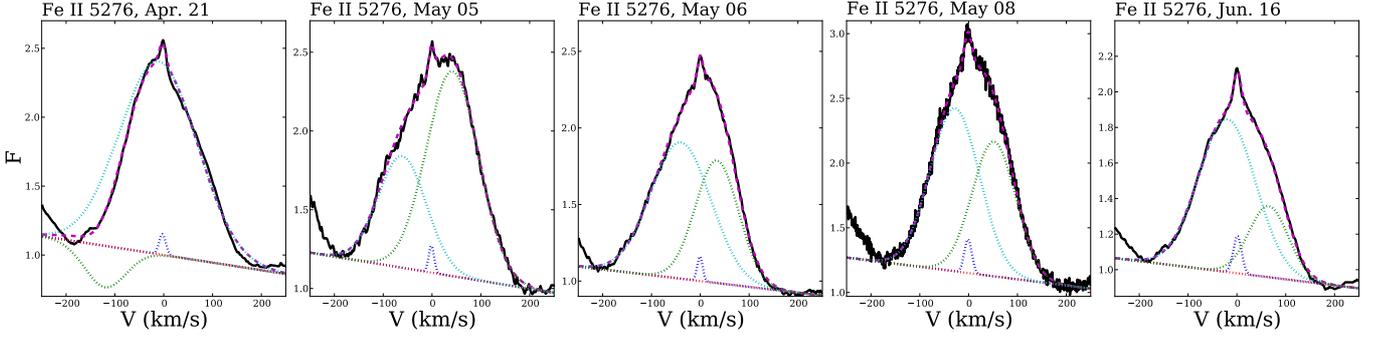}
\caption{Fe~II 5276\AA\ line: Results of the fit with three Gaussian components
for the different outburst epochs. The original data
is shown as the bold line, individual Gaussians
are marked by dotted lines, and the final fit is shown as a dashed line.   \label{FeII5276fit-fig}}
\end{figure*}

The observations of 2008 April 21 and June 16 are harder to interpret.
While the NC remains stable, the BC profiles can be reproduced in 
several ways. The ionized lines can be well-described in terms of a broad
emission component with blueshifted and/or redshifted absorption.
There is a tendency for the April 21 data to be best fitted with a blueshifted
absorption (indicative of wind), while a redshifted absorption most closely fits the
June 16 data, although most spectra have signs of both components. The
distinction is not so clear in the neutral metallic lines, which can be fitted with
similar wind/infall components, but also with combinations of two Gaussian components
in emission, although in these cases the shifts between the Gaussian
components are less extreme than in May. Despite the more evident signatures of wind and infall
in the ionized metallic lines, we are also unable to fully rule out the presence of two different
Gaussian components in the line profiles of the ionized lines.

In summary, the rapid variations observed in May are indicative of a rapid dynamical change, probably caused
by rotation plus infall of gas clumps, and the simple line-fitting confirms that the observed metallic lines
are indeed consistent with a non-axisymmetric structure with several distinct components.
The models of optical photometry and mid-IR interferometric observations 
during the outburst phase developed by Juh\'{a}sz et al. (2012) 
proposed that a hot component dominated the 
spectral energy distribution (SED) at
optical wavelengths, which could be assimilated to a large hot spot
on the stellar surface, or a narrow ring of hot gas at the inner
edge of the dusty disk. The net velocity shifts observed in the optical 
lines agree with the Keplerian velocity of material near the
inner edge of the dusty disk (0.2-0.35 AU; Sipos et al. 2009), although 
the presence of redshifted absorption features is a clear sign of infall,
so we have most likely a combination of rotation and infall.
The rapid variations in the line profiles and velocities suggest that,
rather than an axisymmetric hot, rotating disk (or ring), we 
observe a non-axisymmetric
structure that spirals in onto the star from distances consistent with the inner disk rim. 
This agrees with the CO observations by
Goto et al. (2011) and the near-IR data by K\'{o}spal et al. (2011), 
which detected an asymmetric structure
that could correspond to either an accretion column or matter flow spiraling
in onto the star, although the CO emission is restricted to a cooler
region than the source of the metallic lines. Rotation and infall of a non-axisymmetric,
thick accretion flow would naturally shift the bulk velocity
of the material as observed in the spectra. On the other hand, the 
fact that the broad component is always visible places limits on the
possibility of occultation by the star, and suggests that either we
view the system (or the accretion flow) from a high angle, or the
structure, although not axisymmetric, is relatively extended in
the azimuthal and/or vertical
direction. This includes the possibility of having more than one accretion column,
and is also consistent with the astrometric signal detected by K\'{o}spal et al.(2011)
for hydrogen emission. The parcels of gas traveling within the 
column would increase in temperature
and velocity as they approached the star. A similar scenario, although with
cold gas clumps instead of hot matter, can explain the transient absorption
events seen in UXors and Herbig Ae stars (Mora et al. 2002, 2004), although our
more scarce data does not allow us to define a rotation pattern.

\subsection{Physical conditions in the line-emitting region \label{int}}

The relative intensities of the emission lines can be used to investigate
the physical conditions (in particular of density and temperature)
of the regions where the emission originates, and whether the 
emitting region can be considered to be in local thermodynamical 
equilibrium (LTE). The line ratios for many lines commonly seen in CTTS have 
been calibrated in terms of temperature and density, and can also
provide information about saturation or collisional de-excitation. For other lines,
the Saha equation can be applied to derive the physical properties of the 
emitting gas, assuming LTE. 
Here, we combine these methods to constrain the physical
conditions in the line-emitting region.

The accretion-related emission lines in EX Lupi display typical CTTS
behavior in both the quiescence and the outburst phases.
In the H~I Balmer series (especially, H$\alpha$ and
H$\beta$), we observe the usual
saturation pattern found in CTTS (e.g. Hamann \& Persson 1992a). The
lower Balmer lines are easily saturated under the typical conditions
in CTTS.
The components of the Ca~II IR triplet\footnote{The 8542\AA\ component
is affected by the FEROS gap, but the intensity of the line during the outburst can be estimated
from the observed peak, even if the line in full is not visible.} have relative intensities
closer to unity than to the 1:9:5 relation expected from their
transition coefficients, which is suggestive of formation in
high density environments, as is also common in CTTS (Hamann \& Persson 1992a). 
This sets a lower limit to the electronic density of 10$^{11}$ cm$^{-3}$
for temperatures in the range 4000-10000 K (Hamann \& Persson 1992a).
The Na I doublet usually appears to be close to saturation in CTTSs, and
this is also consistent with what we observe in EX Lupi, although instead
of the normal 2:1 ratio, we observe instead that the strength of the second component is
about $\sim$80\% of the first one. During the quiescence phase, 
the emission lines are rarely blended. The Mg I 
lines at 5167\AA, 5172\AA, and 5183\AA\ show similar intensities in the pre- and post-outburst
spectra, as expected from LTE. The O~I triplet at 7771/4/5\AA\
also displays intensities consistent with the theoretical LTE values in quiescence.
Similar estimates in outburst are highly uncertain owing to the uncertainty in the
continuum levels and the presence of blends.
We can also provide some initial constraints on the temperature by
considering an upper limit of 10000 K (to ensure the presence of neutral hydrogen)
and even $<$7000 K, applying the requirement that the metallic lines and the H~I
emission are produced in different places, as suggested by their different dynamics
(Beristain et al. 1998).
A lower limit to the temperature can be set as the temperature of 
an M0 star ($\sim$3800 K).
Juh\'{a}sz et al.(2012) found that the optical SED in outburst 
is dominated by a continuum
emission that can be fitted as black body emission of temperature $\sim$6500 K,
which is also a good indicator of the dominant temperatures in the line-emitting region
during the outburst. 

We can check both the temperature and density conditions of
the emitting region by using the Saha equation (Mihalas 1978) for a two-component
atom-ion pair. This provides only a limited 
approach: in the real case, we would have to include a large number
of atoms and ions of different species and their levels, and consider that 
the LTE assumption may not be correct. Nevertheless, 
this simple exercise can provide some basic information on the order
of magnitude of temperatures and densities required in the line formation
region. The Saha equation gives the relation
between two successive states of ionization of a given species in LTE

\begin{equation}
	\frac{N_{j+1}n_e}{N_j} 
	= 
	\left(\frac{2\pi m k T}{h^2}\right)^{3/2}
	\frac{2 U_{j+1}(T)}{U_j(T)}
	e^{-\chi_I / kT},
\label{eq:2}
\end{equation}
where, N$_{j+1}$, N$_j$, and n$_e$ represent the number of atoms in the j+1 and j
ionization states and the electron number density, T is the temperature, m is the electron
mass, $\chi_I$ is
the ionization potential, and U$_{j+1}$ and U$_j$ are the partition functions for the
j+1 and j states.  The level populations are calculated according to the
Boltzmann distribution, and can thus be transformed into expected
line intensity ratios to compare with the observed data.
For our estimates, we used the data and Saha equation calculator
provided by the NIST database, varying the electron density and temperature
within reasonable limits and checking a single pair atom/ion at a time.

For the outburst, we first concentrated on 
the very abundant Fe~I and Fe~II lines, estimating the approximate ratios
from line peaks. This is only valid if the line profiles are 
similar and we can assume that both Fe~I and Fe~II originate in the
same region (we discuss the origin perhaps not
being exactly the same later on). 
In general, for a given wavelength, Fe~II lines are twice
as strong as Fe~I lines  (Figures \ref{FeIout-fig} and \ref{FeIIout-fig}). 
To obtain a typical 1:2 line ratio for Fe~I:Fe~II
for a temperature of 6500 K, we need an electronic density of around 
n$_e\sim$5$\times$10$^{12}$ cm$^{-3}$.
This value is consistent with the requirements for the saturation
of the Ca~II IR triplet. The line ratio is very sensitive to the electron
density, so changes by a factor of few would result in
temperatures from 5800 K (for n$_e\sim$1$\times$10$^{12}$ cm$^{-3}$)
up to 7000 K (for n$_e\sim$1$\times$10$^{13}$ cm$^{-3}$).
If we now checked the relative intensities of other neutral/ion pairs
for the same density and temperature ranges, we would expect intensity ratios
of 1-2 orders of magnitude for Ti~II relative to Ti~I lines, of about a factor
of five for Si~II relative to Si~I, and a factor of about two for Cr~II
relative to Cr~I. This is roughly consistent with the observations
(Figure \ref{SiTiCrout-fig}).
In quiescence, this exercise becomes harder owing to the lower S/N and the smaller 
number of observed lines. When we concentrate again on the Fe I/II lines, we
observe similar or slightly larger Fe II:Fe I ratios (2-3). Since the
stronger lines are those with lower excitation potentials, the temperature is
expected to be lower than in the outburst phase, so the differences in the line
ratios are probably caused by a lower electron density.

For the He~I/He~II lines, if we assumed the minimum electronic density 
n$_e\sim$10$^{11}$cm$^{-3}$ derived from the
Ca~II IR triplet, we would need temperatures in the 15000~K range
to obtain a detectable emission of He~II at 4686\AA. Since He~II is
observed in both outburst and quiescence, this is valid for all epochs. Higher
densities would require even hotter temperatures, but too high electron
densities would produce He~I emission lines at
longer wavelengths (which are not
detected) comparable or stronger than the detected He~II line. We would not
expect any detectable He  emission assuming the temperature ranges
required for the metallic lines, but the line profiles clearly have
different origins. The He~I NC and the He~II line
are most likely produced in the transition zone between the chromosphere and
the stellar corona, at high densities and temperatures. The He~I
BCs, on the other hand, probably originate in the hottest parts
of the accretion flow.
To produce the H~I Balmer series in emission we typically need lower temperatures than
for the He lines, and the differences in the profiles of H I and He I lines
imply that they are formed in different places. We can explain why the Balmer 
lines do not show any evidence of
the dynamical effects seen in the metal lines
considering that the H I emission of the
metallic-emitting region is overwhelmed by the H I emission
from hotter environments closer to the star. 

The NC and BC of the metallic lines in the outburst phase are produced
in different environments, since the superposition of the physical
locations of the NC and BC areas would result in the BC being the
only visible one. Estimating the line ratios for the NC would require 
more accurate line models,
given the difficulties in extracting the line from a profile-variable BC and
an unknown continuum level. In principle, since all metallic lines
have a NC+BC structure, and since we do not see any NC standing
alone (although in some very faint Ti~I lines, the BC is
undetected among the strong nearby lines and continuum), we 
would expect similar temperatures in the region responsible
for the NC. A temperature of $\sim$6500 K would be consistent with 
a chromospheric origin.
The enhanced Ti~I NC emission could be reproduced assuming either slightly
lower temperatures or higher electron densities.
The NC was stronger in June,
which is probably due to the decrease in the continuum
and BC emission related to the lower accretion rates.

An independent density analysis can be derived from the accretion rate.
If we assume that the line-emitting region we see in the outburst
corresponds to the accretion flow,
the measure of the accretion rates and line velocities can be used to derive
a mass density in the proximities of the star that can be compared with the densities derived by
other means. We consider an accretion rate of 2$\times$10$^{-7}$ M$_\odot$/yr
and the velocities observed in the emitting lines, which have widths of FWHM$\sim$200-300 km/s, depending on
the line, $>$300 km/s for H~I Balmer lines, $\sim$230 km/s for Ca~II lines,
$\sim$180 km/s for typical metallic lines. We note that the line velocities are
very similar to the free-fall velocities expected for magnetospheric infall
from a 5R$_*$ distance (Gullbring et al. 1998), if we assume the stellar
radius to be R$_*$=1.6 R$_\odot$ and a stellar mass of 0.6 M$_\odot$ (Sipos et al. 2009).
Assuming that the full accretion column lands onto
the star, its cross-section can be considered as a fraction 
of the stellar surface, f, with f$<$1. The gas density in the accretion column 
can be estimated from the accretion rate, considering the velocity and 
the cross-section of the flow. Assuming the typical weight of interstellar
matter, $\mu$=1.36$\times$10$^{-24}$ g, we arrive at a number density
of $\sim$1.3/f $\times$10$^{12}$ cm$^{-3}$. If the coverage fraction were
on the order of 10\% of the stellar surface, the density would thus
be $\sim$10$^{13}$ cm$^{-3}$.  For the
relevant temperature ranges, we can assume that all metals are essentially ionized but
that hydrogen and helium remain neutral, so that
the electron density is very similar to the number density of metals, or
about 0.1\% of the total particle density, assuming a solar composition for
the accretion flow.
Nevertheless, this would result in an electron density two to three
orders of magnitude below the estimate for a temperature $\sim$6500 K, and
one to two orders of magnitude below the requirement for saturation 
in the Ca~II IR triplet. Therefore, the most plausible
solution is that we have strong extra ionization, probably due to energetic radiation
from the star, which ionizes up to a few percent of the H (see
for instance Kwan \& Fischer 2011).
Another solution would be to assume that the emitting region is enhanced in
metals,
but this would be hard to achieve if we identify the origin of the
metallic emission with an accretion column, considering that
the increased accretion event observed in 2008 resulted in
a total mass accretion of about 50-70 Earth masses\footnote{ Note that some degree of metal enrichment or depletion,
for instance, caused by deviations in the standard dust-to-gas ratio, is possible due
to grain growth, photoevaporation, and the potential formation of protoplanets, 
which are all processes expected to occur to some degree in typical circumstellar disks.}.

To summarize, assuming reasonable densities
(n$_e\sim$10$^{11}$-10$^{16}$cm$^{-3}$) and temperatures (T$\sim$4000-10000 K)
for an accretion flow environment,
we obtain a minimum temperature of 4500 K
to attain the observed Fe~II:Fe~I line intensity ratio of $\sim$2:1 with
the minimum density required to obtain similar fluxes in the Ca~II IR
triplet during the outburst phase. Choosing the value of $\sim$10000 K, the typical temperature in
the accretion shock of T Tauri stars (Hartigan et al. 1990; Gullbring et al. 1998), 
results in densities as high as 10$^{16}$cm$^{-3}$ to ensure that there is a large enough Fe~I 
fraction, but this temperature seems excessive for Ti~I emission, unless
we consider much higher densities, which are hard to attain with the measured 
accretion rates (at least, over large regions with a large velocity spread).
 A high temperature ($>$7000 K) is also inconsistent with the lack of 
variable dynamical signatures in the H~I lines, relative to the metallic lines.
The lower E$_k$ observed in the quiescence phase point to lower temperatures,
so that the larger Fe II: Fe I ratio observed in quiescence would require lower
electron densities.

As a last remark, we need to consider that the emitting region has most likely a non-uniform
distribution of temperatures and densities. That the line profiles
of neutral atoms and ions during the outburst phase are different, with the neutral atoms being more
sensitive to the velocity variations in the accretion column and the
ions displaying stronger signatures of wind and infall, suggests that
they have slightly different origins, or that the predominant emission
region is not the same in both cases. Therefore, 
the BC of the Fe~II and other ionized lines are probably produced in a less
dense region 
(for instance, the outer parts of the accretion column), while the bulk of
the neutral emission arises from a denser region in the hot, non-axisymmetric
accretion flow. Evaporation in the outer layers of the accretion flow could
also explain why the wind-related blueshifted signatures are more
pronounced in the ionized lines. A more detailed calculation of temperature and
density would require complex models including the accreting star, its
active chromosphere and hot spots, the non-axisymmetric accretion flow,
and a disk wind component.

\subsection{Dynamics and physical conditions in the outburst phase \label{dynphys}}

The excitation potentials (E$_k$) and transition probabilities (or Einstein
coefficients, A$_{ki}$) of the lines can be roughly
translated into temperatures and densities. Transitions with higher
E$_k$ typically require a higher temperature, and those with lower Einstein
coefficients (longer transition times) tend to originate in regions of lower
density. 
As we have mentioned before, in quiescence
the lines with lower excitation potentials (E$_k\sim$2-4 eV) tend to be 
stronger than those with higher excitation potentials (E$_k\sim$4-6 eV), but the situation is
the opposite during the outburst phase: lines with lower excitation potentials
tend to be weaker than those with 
higher excitation potentials. Lehmann et al. (1995) reported the inverse
effect during the 1993-4 EX Lupi outburst, with the low excitation potential
lines being stronger at outburst, which could be interpreted as  cooling
of the region where the lines are produced.
In contrast, our observations suggests that the line-emitting
zone was hotter and more extended in outburst than in quiescence,
which could be explained by the increase in heating caused by the enhanced accretion.

%\begin{landscape}
\begin{table*}
\caption{Significance of the potential correlations between line dynamics, E$_k$, and A$_{ki}$  } 
\label{test-table}
\begin{tabular}{l c c c c c l}
\hline\hline
Quantities & R/p (1) &  R/p (2) &  R/p (3) & R/p (4) & R/p (5) & Comments\\ 
\hline
Neutrals: V vs. E$_k$ & 0.14/0.49 & -0.41/0.03 & -0.26/0.19 & 0.07/0.76 & 0.25/0.21 & No correlation \\
Ions: V vs. E$_k$ &  -0.13/0.47 & -0.10/0.58 & -0.37/0.04 & -0.32/0.11 & -0.26/0.15 & No correlation \\
Neutrals: FWHM vs. E$_k$ & 0.47/0.01 & 0.53/0.004 & 0.49/0.01 & 0.48/0.02 & 0.25/0.21 & Correlation (except on 5) \\
Ions: FWHM vs. E$_k$ & 0.07/0.73 & 0.06/0.74 & 0.11/0.56 & 0.25/0.22 & -0.06/0.75 & No correlation \\
Neutrals: V vs. A$_{ki}$ & -0.19/0.34 & -0.20/0.32 & -0.21/0.30 & 0.09/0.70 & 0.04/0.84 & No correlation \\ 
Ions: V vs. A$_{ki}$ & -0.39/0.03 & -0.55/0.001 & -0.36/0.05 & -0.21/0.30 & -0.30/0.11 & Mild anticorrelation \\
Neutrals: FWHM vs. A$_{ki}$  & 0.07/0.74 & 0.17/0.40 & 0.06/0.78 & 0.51/0.02 & -0.08/0.78 & No correlation \\
Ions: FWHM vs. A$_{ki}$ &  0.56/0.001 & 0.57/9E-4 & 0.44/0.01 & 0.43/0.03 & 0.61/3E-4 & Correlation \\ 
\hline
\end{tabular}
\tablefoot{Potential correlations between the line dynamics (represented as center velocity [V]
and FWHM of the BC of the metallic neutral and ionized lines) and physical conditions 
(represented by E$_k$ and A$_{ki}$) during the outburst phase.
The results of a Spearman rank test are tabulated for each line type (ions/neutrals) and date, 
with R being the correlation coefficient, and p the false alarm probability. (1) 2008 April 21;
(2) 2008 May 05; (3) 2008 May 06; (4) 2008 May 08; (5) 2008 June 16.}
\end{table*}
%\end{landscape}

We have thus explored the potential correlation of E$_k$ and A$_{ki}$
with the line properties, given by their central line velocity and FWHM,
of neutral and ionized metallic lines observed during the outburst (Table \ref{profi-table}). 
There is no evident correlation of E$_k$ with the central line velocity of
the neutral and ion lines (Table \ref{test-table}). A Spearman test gives probabilities that 
the two quantities are uncorrelated of 4-60\% for ions, and 3-75\% for
neutrals. It would seem that the correlation is sometimes significant, but 
that the correlation coefficients change sign over the different days indicates
that both quantities are uncorrelated. On the other hand, there is a mild correlation
of the FWHM of the neutral atoms with E$_k$ in all but the last observation, 
with positive correlation coefficients of
around 0.5, and probabilities of 0.4-2\% that the two quantities are uncorrelated.
This would be consistent with a larger FWHM being related to
higher turbulence, and thus higher temperature regions, where the lines
with higher  E$_k$ can be excited. There is no evident correlation between FWHM and E$_k$ 
for ions.
Regarding A$_{ki}$, there is a weak anticorrelation between the central line velocity
of the ionized lines and A$_{ki}$, which would be consistent with the higher velocity components coming from
the less dense regions, as expected in a wind scenario. The Spearman test indicates that the probability that these
two quantities are uncorrelated are 3\%, 0.1\%, and 5\% for the first three observations,
but the data are uncorrelated in the last two cases. At the same time, there is a clear correlation
between the FWHM of ions and A$_{ki}$ (with Spearman probabilities in the range 0.03-1\% that the two quantities
are uncorrelated), which can be understood as a correlation between the higher temperatures (larger
FWHM) and higher densities (large A$_{ki}$). No such correlations are observed for
the neutral lines. 

In summary, we find evidence that the lines with larger FWHM originate in hotter, denser
environments, and that the ionized lines are dominated by less dense regions. This
is consistent wiht formation in an infalling accretion column plus wind,
but the general lack of strong trends points to 
a complex line-forming structure.

\section{Conclusions \label{conclu}}
We have presented a collection of optical spectra of EX Lupi taken before,
during, and after the 2008 outburst. The spectra contain
a large number of emission lines. There is strong emission from the typical, accretion-related
lines commonly observed in CTTS (H~I Balmer and Paschen series, Ca~II, He~I).
In addition, the spectra are rich in neutral metallic emission lines (Fe~I, Ca~I,
Cr~I, Ni~I, Co~I, Mg~I, K~I, C~I), which
appear superimposed on the photospheric absorption lines in the quiescence spectra, and 
an abundant range of metallic ionized lines (Fe~II, Si~II, Ti~II, Cr~II).
The emission lines overwhelm any absorption feature during the outburst phase. 
We analyze these emission lines in order to determine the physical regions,
conditions, and dynamics involved
in the accretion processes, and how they changed during the phases of
increased accretion in 2008.

\begin{figure}
   \centering
   \resizebox{\hsize}{!}{\includegraphics{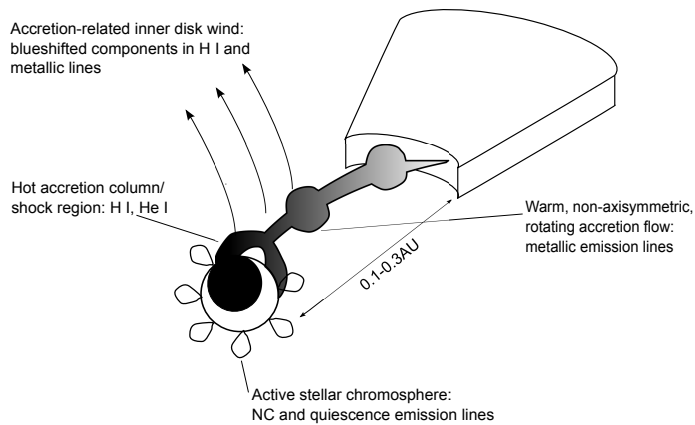}}
\caption{ A sketch of the proposed morphology of the accretion flow in
EX Lupi to explain the optical spectroscopy data (not to scale). One or more clumpy
accreting structures could be present in the system, and the geometry of the
inner disk, accretion column(s), and accretion-related wind is unconstrained. \label{exlupisketch-fig}}
\end{figure}

   \begin{enumerate}
\item The data are consistent with 
variable accretion being responsible for the observed outburst, with an increase 
of up to 2-3 orders of magnitude between quiescence
and outburst. A
wind (most likely an accretion-related inner disk wind) 
also developed in outburst, with an intensity that was correlated with the
accretion rate that decreased as the outburst
faded. The wind covers a large range of velocities, from -50 to -200 km/s. 
The presence of this wind agrees with 
the independent results of Goto et al. (2011) and
K\'{o}sp\'{a}l et al. (2011). During the quiescence phases, wind signatures were either
weak or absent, and the main feature in the H I Balmer lines was a redshifted 
absorption consistent with infall. 
The development of a strong wind in the outburst maximum is exceptional in
the observed EX Lupi outbursts, and a sign of an accretion rate higher than
in typical EXor outbursts, although still well below the values seen in FUors.
Despite the strong wind, 
there is no evidence of any forbidden lines in emission related to shocks, so the 
presence of remnant envelopes or nearby molecular cloud material can be excluded.

\item The similarity between the pre- and post-outburst spectra is indicative of
the rapid recovery of the system, suggesting that there was no
dramatic change in the structure of the inner gaseous disk and accretion columns after
the episode of increased accretion. The outburst is thus consistent with 
a strong variation in the accretion rate through remarkably stable 
accretion channels. The outburst represents thus an ideal environment to
explore the way accretion proceeds in this object, and maybe also in
similar EXors and even typical CTTS.

\item The observed metallic emission lines are narrow in quiescence and
have a NC and a BC during the outburst. Their excitation potentials are consistent
with the temperatures expected in the stellar chromosphere, hot spots, accretion columns, and
shock region.
Both the NC and the narrow emission lines observed in quiescence are consistent
with an origin in an active chromosphere.
The velocities observed in the BC during the outburst are too large for an origin
near the stellar surface, and are thus most likely produced in an 
accretion column or a similarly extended and non-axisymmetric
structure that heated up during the increased accretion episode.

\item We have observed dynamical changes in the BC of the metallic lines, that have
timescales of days and are consistent with movement of large parcels
of gas. The velocity variations were maximal on 2008 May 5, when the
BC developed a secondary redshifted peak at +50 km/s, which became 
much fainter 24h later, and absent after 72h. This is
consistent with the dynamics of a
hot and non-axisymmetric accretion flow or column(s), where independent
parcels of accreting gas spiral-in in a clumpy accretion scenario
(see a sketch of the system in Figure \ref{exlupisketch-fig}). 
The picture of a rotating/infalling accretion flow is
consistent with the observations of Aspin et al. (2010), 
the CO models of Goto et al.(2011; although note that CO traces a
cooler region within the accretion flow/inner disk), and the interferometric
observations of Juh\'{a}sz et al. (2012), which detected 
non-axisymmetric gaseous structures within the innermost disk. 
In addition, examples of
the clumpy infall of gas parcels have been observed in UXor and
Herbig Ae stars (Mora et al. 2002, 2004), although in these
cases the infalling matter is colder and produces absorption
features.  The lack of dynamical signatures in the H~I lines, compared
to the metallic lines, can be explained if the H~I emission is dominated by
a distinct, hotter environment, compared to the metallic emission.

\item Juh\'{a}sz et al. (2012) found the optical component of the SED in 
outburst to
be dominated by emission from a region with temperature $\sim$6500 K. For such
a temperature, if we assumed LTE, we would need a density of a few times 
10$^{12}$cm$^{-3}$ in the line emitting region in order to reproduce
the observed line ratios for neutral/ion pairs (Fe~I/Fe~II, Si~I/Si~II, Ti~I/Ti~II,
and Cr~I/Cr~II), which would also ensure the saturation of the
Ca II IR triplet. Nevertheless, the line profiles suggest 
that the ionized lines may originate in a slightly
different environment from the neutral lines. The
stronger signatures of winds observed for ionized lines point to
an origin in less dense
areas dominated by the accretion-powered wind and/or by evaporation in the surface 
of the accretion column. In this case, the neutral region of the accretion flow could 
have a lower temperature and/or higher density.
Extra sources of ionization (probably, the energetic radiation of the star) are
required to achieve the desired electronic densities in the accretion column.
   \end{enumerate}

\begin{acknowledgements}
We thank the referee for a report that contributed to clarify this paper.
A.S.-A. acknowledges support by the Deutsche Forschungsgemeinschaft, grant number
SI 1486/1-1, and the Spanish ``Ram\'{o}n y Cajal" program, funded by the MICINN.
We acknowledge the use of EX Lupi observations from the AAVSO 
International Database contributed by observers worldwide.\end{acknowledgements}

%\clearpage
\onecolumn

\longtabL{4}{
\begin{landscape}
\begin{longtable}{lccccccccccccl}
\caption{\label{profi-table} Independent lines observed during the outburst phase.
(1) JD4577 (2) JD4591 (3) JD4592 (4) JD4594 (5) JD4633. 
Note that the HARPS observations (4) have a smaller wavelength coverage,
resulting in several lines not being observed.
`M:' marks lines that might be merged with other weaker, nearby lines.} \\
\hline\hline
 Line $\lambda_{lab}$ &E$_k$  &A$_{ki}$  &$\lambda_{obs}$(1) & EW(1) / F(1)&$\lambda_{obs}$(2) & EW(2) / F(2) &$\lambda_{obs}$(3) & EW(3) / F(3) &$\lambda_{obs}$(4) & EW(4) / F(4) &$\lambda_{obs}$(5) & EW(5) / F(5) & Comments\\ 
  (\AA) & (eV) & (s$^{-1}$) & (\AA) & (\AA / km/s) & (\AA) & (\AA / km/s) & (\AA) & (\AA / km/s) & (\AA) & (\AA / km/s) & (\AA) & (\AA / km/s) & \\ 
\hline
\endfirsthead
\caption{Continued.}\\
\hline\hline
 Line $\lambda_{lab}$ &E$_k$  &A$_{ki}$  &$\lambda_{obs}$(1) & EW(1) / F(1)&$\lambda_{obs}$(2) & EW(2) / F(2) &$\lambda_{obs}$(3) & EW(3) / F(3) &$\lambda_{obs}$(4) & EW(4) / F(4) &$\lambda_{obs}$(5) & EW(5) / F(5) & Comments\\ 
  (\AA) & (eV) & (s$^{-1}$) & (\AA) & (\AA / km/s) & (\AA) & (\AA / km/s) & (\AA) & (\AA / km/s) & (\AA) & (\AA / km/s) & (\AA) & (\AA / km/s) & \\ 
\hline
\endhead
\hline
H~I 4861.28 & 12.75 & 1.72e+07 & 4860.96 & -24.8 / 343 &  4861.01 & -22.6 / 359 &  4860.99 & -24.6 / 339 &  4861.27 & -22.1 / 396 &  4861.12 & -17.7 / 344 & \\  
H~I 6562.57 & 12.09 & 5.39e+07 & 6562.56 & -61.0 / 278 &  6562.43 & -78.0 / 303 &  6562.35 & -84.0 / 295 &  6562.61 & -71.5 / 321 &  6562.66 & -67.3 / 310 & \\  
H~I 8598.39 & 13.53 & 9.21e+03 & 8598.49 & -4.3 / 210 &  8598.67 & -3.6 / 208 &  8598.03 & -3.4 / 214 &  --- & --- / --- &  8598.39 & -3.0 / 185 & \\  
H~I 8750.46 & 13.54 & 2.02e+04 & 8750.91 & -6.0 / 210 &  8751.03 & -5.2 / 211 &  8750.37 & -4.7 / 199 &  --- & --- / --- &  8750.81 & -5.5 / 220 & \\  
He~I 5875.60 & 23.07 & 5.30e+07 & 5875.78 & -4.1 / 199 &  5875.83 & -2.7 / 178 &  5875.57 & -2.8 / 181 &  5876.03 & -3.9 / 196 &  5875.84 & -2.4 / 164 & \\  
He~I 6678.151 & 23.07 & 6.37e+07 & 6678.31 & -2.9 / 168 &  6678.62 & -2.2 / 160 &  6678.17 & -2.2 / 158 &  6678.44 & -3.5 / 173 &  6678.31 & -2.1 / 156 & \\  
He~I 7065.19 & 22.72 & 9.28e+06 & 7065.62 & -1.7 / 211 &  7065.74 & -1.4 / 250 &  7065.56 & -1.3 / 207 &  --- & --- / --- &  7065.67 & -1.3 / 197 & \\  
Mg~I 4702.991 & 6.98 & 2.19e+07 & 4703.04 & -0.7 / 149 &  4703.43 & -0.8 / 169 &  4702.92 & -1.1 / 191 &  4703.25 & -1.6 / 170 &  4703.04 & -0.7 / 149 & \\  
Mg~I 8806.756 & 5.75 & 1.27e+07 & 8806.90 & -7.5 / 167 &  8807.37 & -7.6 / 178 &  8806.77 & -6.6 / 162 &  --- & --- / --- &  8806.92 & -7.1 / 176 & \\  
Si~II 6347.10 & 10.07 & 5.84e+07 & 6346.96 & -2.8 / 213 &  6346.93 & -2.7 / 234 &  6346.55 & -1.7 / 199 &  6347.09 & -3.0 / 194 &  6346.59 & -2.2 / 233 & M: \\ 
Si~II 6371.36 & 10.07 & 6.80e+07 & 6370.56 & -2.6 / 217 &  6370.56 & -2.5 / 228 &  6369.91 & -1.8 / 188 &  6370.66 & -2.7 / 199 &  6370.27 & -2.0 / 215 & M: \\ 
Ca~I 6102.72 & 3.91 & 9.6e+06 & 6102.84 & -0.9 / 161 &  6103.40 & -0.8 / 152 &  6102.78 & -0.8 / 167 &  6103.01 & -1.3 / 163 &  6102.86 & -0.8 / 168 & \\  
Ca~I 6122.22 & 3.91 & 2.87e+07 & 6122.17 & -0.8 / 158 &  6122.68 & -1.0 / 164 &  6122.07 & -0.8 / 161 &  6122.41 & -1.5 / 166 &  6122.17 & -0.8 / 158 & \\  
Ca~I 6717.69 & 4.55 & 1.2e+07 & 6717.71 & -0.6 / 172 &  6718.42 & -0.5 / 142 &  6717.56 & -0.4 / 160 &  6717.81 & -0.9 / 179 &  6717.57 & -0.5 / 156 & \\  
Ca~II 8498.02 & 3.15 & 1.11e+06 & 8497.84 & -70.8 / 213 &  8498.15 & -79.4 / 229 &  8498.12 & -80.1 / 211 &  --- & --- / --- &  8498.18 & -64.6 / 211 & \\  
Ca~II 8662.14 & 3.12 & 1.06e+07 & 8661.74 & -63.7 / 248 &  8662.02 & -68.4 / 258 &  8662.07 & -67.2 / 242 &  --- & --- / --- &  8662.17 & -58.0 / 246 & \\  
Ca~II 8912.07 & -100 & -100 & 8912.30 & -3.0 / 161 &  8912.67 & -2.6 / 178 &  8911.92 & -2.1 / 161 &  --- & --- / --- &  8912.23 & -2.5 / 156 & \\  
Ca~II 8927.36 & -100 & -100 & 8927.39 & -3.8 / 166 &  8927.51 & -2.8 / 185 &  8926.90 & -2.5 / 168 &  --- & --- / --- &  8927.20 & -2.8 / 164 & \\  
Ti~II 4330.695 & 4.04 & 8.1e+05 & 4330.81 & -1.6 / 178 &  4331.06 & -1.7 / 163 &  4330.73 & -1.3 / 153 &  4330.85 & -2.3 / 166 &  4330.81 & -1.6 / 178 & \\  
Ti~II 4563.77 & 3.94 & 8.8e+06 & 4563.98 & -2.3 / 182 &  4564.07 & -2.9 / 198 &  4563.82 & -2.9 / 188 &  4564.11 & -3.6 / 179 &  4563.98 & -2.3 / 182 & \\  
Ti~II 4571.98 & 4.28 & 1.2e+07 & 4571.88 & -3.6 / 194 &  4571.94 & -4.8 / 212 &  4571.71 & -4.4 / 182 &  4572.01 & -5.7 / 186 &  4571.88 & -3.6 / 194 & M: \\ 
Ti~II 4629.279 & 3.86 & 2.2e+05 & 4629.46 & -4.8 / 171 &  4629.61 & -6.4 / 187 &  4629.34 & -6.3 / 166 &  4629.55 & -7.1 / 158 &  4629.46 & -4.8 / 171 & \\  
Ti~II 4798.521 & 3.66 & 1.8e+05 & 4798.90 & -1.0 / 181 &  4799.18 & -0.9 / 173 &  4798.67 & -0.8 / 159 &  4798.94 & -1.5 / 183 &  4798.84 & -1.1 / 201 & M: \\ 
Ti~II 4805.10 & 4.64 & 1.1e+07 & 4805.20 & -2.0 / 182 &  4805.34 & -1.7 / 181 &  4805.01 & -1.5 / 168 &  4805.26 & -2.0 / 155 &  4805.09 & -1.4 / 175 & \\  
Ti~II 4911.193 & 5.65 & 3.2e+07 & 4911.18 & -1.8 / 186 &  4911.42 & -1.8 / 203 &  4910.99 & -1.4 / 172 &  4911.26 & -2.3 / 184 &  4911.12 & -1.3 / 186 & \\  
Ti~II 5226.56 & 3.94 & 3.1e+06 & 5226.84 & -4.6 / 183 &  5227.07 & -4.4 / 191 &  5226.77 & -4.1 / 182 &  5226.98 & -5.4 / 179 &  5226.87 & -3.3 / 183 & \\  
Cr~II 4824.127 & 6.44 & 1.7e+06 & 4824.24 & -3.4 / 180 &  4824.43 & -2.8 / 182 &  4824.07 & -2.5 / 171 &  4824.33 & -3.7 / 167 &  4824.19 & -2.3 / 176 & \\  
Cr~II 4848.235 & 6.42 & 2.6e+06 & 4848.51 & -2.2 / 171 &  4848.71 & -2.2 / 184 &  4848.37 & -1.7 / 161 &  4848.57 & -2.7 / 163 &  4848.45 & -1.5 / 161 & \\  
Fe~I 3872.501 & 4.19 & 1.05e+07 & 3872.73 & -1.9 / 183 &  3872.89 & -1.8 / 170 &  3872.66 & -1.9 / 195 &  3872.86 & -2.4 / 187 &  3872.74 & -1.8 / 198 & \\  
Fe~I 4045.594 & 6.27 & 7.39e+06 & 4045.51 & -1.7 / 165 &  4045.68 & -2.3 / 215 &  4045.54 & -2.0 / 182 &  4045.76 & -2.6 / 183 &  4045.63 & -1.3 / 172 & \\  
Fe~I 4191.43 & 5.43 & 2.73e+07 & 4191.23 & -0.8 / 178 &  4191.52 & -0.9 / 183 &  4191.28 & -0.7 / 152 &  4191.37 & -1.2 / 171 &  4191.23 & -0.8 / 178 & \\  
Fe~I 4202.029 & 4.43 & 8.22e+06 & 4202.23 & -0.9 / 146 &  4202.31 & -2.0 / 192 &  4202.14 & -1.4 / 148 &  4202.31 & -2.6 / 173 &  4202.19 & -1.0 / 156 & \\  
Fe~I 4957.597 & 5.31 & 1.18e+07 & 4957.57 & -1.9 / 157 &  4957.85 & -2.4 / 169 &  4957.57 & -1.9 / 148 &  4957.68 & -3.0 / 149 &  4957.57 & -1.9 / 157 & \\  
Fe~I 5371.489 & 3.27 & 1.05e+06 & 5371.28 & -3.8 / 184 &  5371.55 & -3.3 / 190 &  5371.30 & -3.1 / 172 &  5371.37 & -4.5 / 183 &  5371.24 & -2.6 / 182 & \\  
Fe~I 5446.874 & 3.88 & 2.50e+04 & 5446.94 & -3.1 / 170 &  5447.27 & -2.6 / 164 &  5446.94 & -2.5 / 155 &  5447.01 & -3.8 / 166 &  5446.93 & -2.2 / 171 & \\  
Fe~I 5455.609 & 3.28 & 6.05e+05 & 5455.64 & -2.6 / 168 &  5456.02 & -2.1 / 158 &  5455.60 & -2.0 / 148 &  5455.74 & -3.0 / 158 &  5455.63 & -1.7 / 155 & \\  
Fe~I 5615.644 & 5.54 & 2.39e+05 & 5615.73 & -1.6 / 162 &  5616.13 & -1.4 / 157 &  5615.61 & -1.3 / 162 &  5615.83 & -2.1 / 164 &  5615.67 & -1.2 / 158 & \\  
Fe~I 5930.18 & 6.74 & -100 & 5929.83 & -0.4 / 185 &  5930.22 & -0.4 / 197 &  5929.63 & -0.4 / 198 &  5930.18 & -0.7 / 200 &  5929.83 & -0.4 / 185 & \\  
Fe~I 6065.482 & 4.65 & 1.07e+06 & 6065.49 & -1.0 / 160 &  6066.05 & -0.9 / 155 &  6065.41 & -0.8 / 156 &  6065.59 & -1.4 / 166 &  6065.44 & -0.7 / 154 & \\  
Fe~I 6191.558 & 4.43 & 7.41e+05 & 6191.68 & -1.9 / 161 &  6192.09 & -1.6 / 145 &  6191.61 & -1.8 / 155 &  6191.66 & -2.3 / 155 &  6191.56 & -1.5 / 166 & \\  
Fe~I 6393.601 & 4.37 & 4.81e+05 & 6393.73 & -1.6 / 162 &  6394.24 & -1.8 / 167 &  6393.63 & -1.6 / 162 &  6393.78 & -2.1 / 163 &  6393.63 & -1.5 / 175 & \\  
Fe~I 6400.001 & 5.54 & 9.27e+06 & 6400.27 & -1.5 / 163 &  6400.72 & -1.3 / 153 &  6400.16 & -1.1 / 150 &  6400.28 & -2.0 / 165 &  6400.12 & -1.2 / 159 & \\  
Fe~I 8824.221 & 3.6 & 3.53e+05 & 8824.38 & -3.9 / 160 &  8825.23 & -3.2 / 137 &  8824.49 & -3.8 / 163 &  --- & --- / --- &  8824.33 & -3.1 / 159 & \\  
Fe~II 4233.167 & 5.51 & 7.22e+05 & 4233.20 & -7.1 / 203 &  4233.32 & -6.1 / 213 &  4233.12 & -5.7 / 190 &  4233.42 & -6.8 / 191 &  4233.27 & -4.7 / 201 & \\  
Fe~II 4351.769 & 5.55 & 4.86e+05 & 4351.83 & -8.1 / 194 &  4351.95 & -6.5 / 201 &  4351.76 & -6.6 / 186 &  4351.97 & -7.8 / 181 &  4351.88 & -5.1 / 189 & \\  
Fe~II 4549.474 & 5.55 & 9.2e+05 & 4549.38 & -7.3 / 212 &  4549.56 & -5.8 / 227 &  4549.36 & -4.7 / 191 &  4549.74 & -6.4 / 198 &  4549.56 & -5.1 / 218 & \\  
Fe~II 5018.434 & 5.36 & 2.0e+06 & 5018.53 & -7.0 / 219 &  5018.38 & -10.3 / 251 &  5018.27 & -10.7 / 222 &  5018.61 & -8.3 / 202 &  5018.53 & -7.0 / 219 & \\  
Fe~II 5234.625 & 5.59 & 2.5e+05 & 5234.73 & -8.9 / 225 &  5234.91 & -8.8 / 243 &  5234.60 & -7.7 / 215 &  5234.90 & -10.7 / 237 &  5234.77 & -6.6 / 232 & \\  
Fe~II 5276.002 & 5.55 & 3.76e+05 & 5275.91 & -8.2 / 205 &  5276.07 & -6.0 / 200 &  5275.83 & -5.3 / 176 &  5276.05 & -6.3 / 172 &  5275.95 & -5.4 / 194 & \\  
Fe~II 5284.109 & 5.24 & 1.9e+04 & 5284.01 & -5.2 / 186 &  5284.24 & -4.6 / 198 &  5283.94 & -4.2 / 173 &  5284.09 & -4.8 / 171 &  5284.00 & -3.5 / 182 & \\  
Fe~II 5316.615 & 5.48 & 3.89e+05 & 5316.65 & -11.2 / 186 &  5316.81 & -9.8 / 201 &  5316.60 & -9.7 / 178 &  --- & --- / --- &  5316.75 & -7.8 / 188 & \\  
Fe~II 5534.847 & 5.48 & 3.0e+04 & 5535.04 & -5.0 / 167 &  5535.33 & -4.6 / 172 &  5534.88 & -4.2 / 157 &  5535.12 & -6.0 / 166 &  5534.99 & -3.4 / 163 & \\  
Fe~II 5991.376 & 5.22 & 4.2e+03 & 5991.51 & -1.6 / 156 &  5991.94 & -1.8 / 167 &  5991.32 & -1.6 / 150 &  5991.56 & -2.0 / 154 &  5991.41 & -1.3 / 153 & \\  
Fe~II 6084.111 & 5.24 & 3.0e+03 & 6084.22 & -0.8 / 155 &  6084.70 & -1.3 / 187 &  6084.06 & -1.1 / 168 &  6084.44 & -1.4 / 168 &  6084.22 & -0.8 / 155 & \\  
Fe~II 6238.375 & 5.88 & 7.5e+04 & 6238.85 & -2.6 / 175 &  6239.23 & -2.4 / 177 &  6238.62 & -3.2 / 199 &  6239.05 & -3.2 / 172 &  6238.79 & -1.8 / 162 & \\  
Fe~II 6247.562 & 5.87 & 1.6e+05 & 6247.48 & -4.3 / 185 &  6247.82 & -3.6 / 184 &  6247.26 & -4.1 / 184 &  6247.60 & -4.9 / 180 &  6247.43 & -2.8 / 172 & \\  
Fe~II 6456.376 & 5.82 & 1.7e+05 & 6456.50 & -5.4 / 160 &  6456.81 & -5.5 / 174 &  6456.36 & -4.8 / 155 &  6456.65 & -6.5 / 158 &  6456.50 & -4.0 / 158 & \\  
Fe~II 6516.053 & 4.79 & 8.3e+03 & 6516.34 & -5.8 / 165 &  6516.63 & -6.2 / 176 &  6516.17 & -5.8 / 157 &  6516.41 & -7.5 / 169 &  6516.30 & -4.5 / 170 & \\  
Fe~II 7462.38 & 5.55 & 2.7e+04 & 7462.60 & -2.6 / 160 &  7463.05 & -2.8 / 169 &  7462.36 & -2.5 / 159 &  --- & --- / --- &  7462.50 & -2.2 / 156 & \\  
Fe~II 7711.71 & 5.51 & 4.94e+04 & 7712.23 & -4.1 / 195 &  7712.63 & -4.2 / 196 &  7711.94 & -3.5 / 183 &  --- & --- / --- &  7712.18 & -3.0 / 179 & \\  
\hline
\end{longtable}
\end{landscape}
}

\Online
\onecolumn

\begin{appendix} %First online appendix
\section{Gaussian fits to the emission lines in outburst}
\begin{figure*}
   \centering
   \includegraphics[width=18cm]{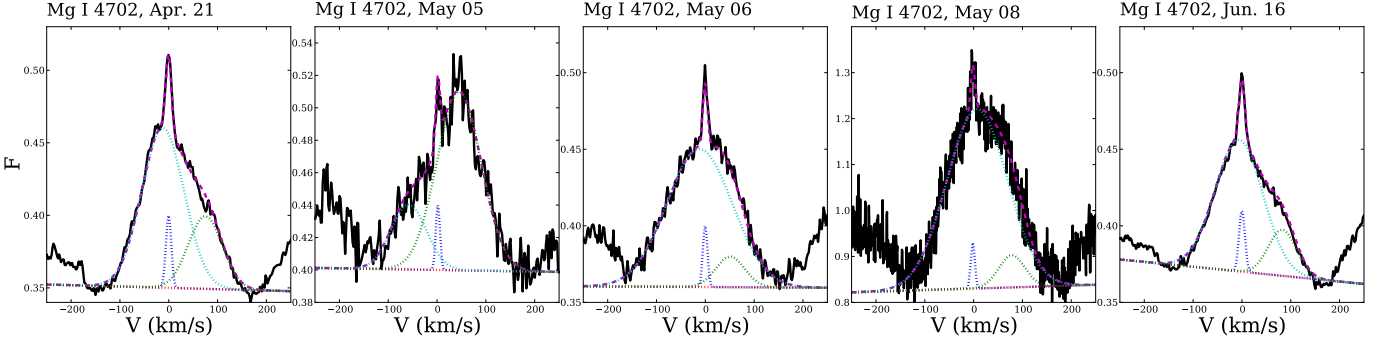}
\caption{Mg~I 4702\AA\ line: Results of the fit with three Gaussian components
for the different outburst epochs. The original data
is shown as the bold line, individual Gaussians
are marked by dotted lines, and the final fit is shown as a dashed line.   \label{MgI4702fit-fig}}
\end{figure*}

\begin{figure*}
   \centering
   \includegraphics[width=18cm]{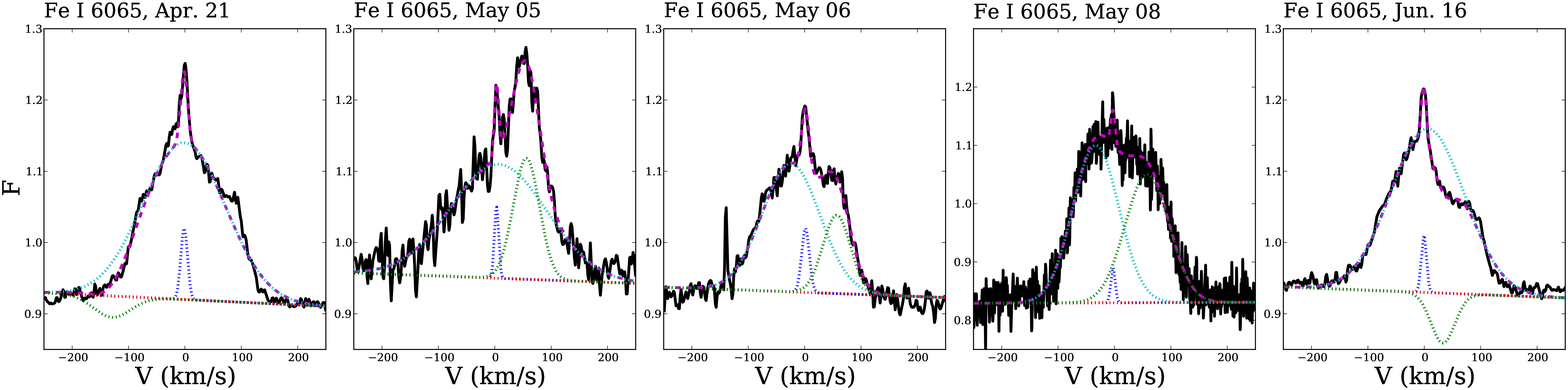}
\caption{Fe~I 6065\AA\ line: Results of the fit with three Gaussian components
for the different outburst epochs. The original data
is shown as the bold line, individual Gaussians
are marked by dotted lines, and the final fit is shown as a dashed line.   \label{FeI6065fit-fig}}
\end{figure*}

\begin{figure*}
   \centering
   \includegraphics[width=18cm]{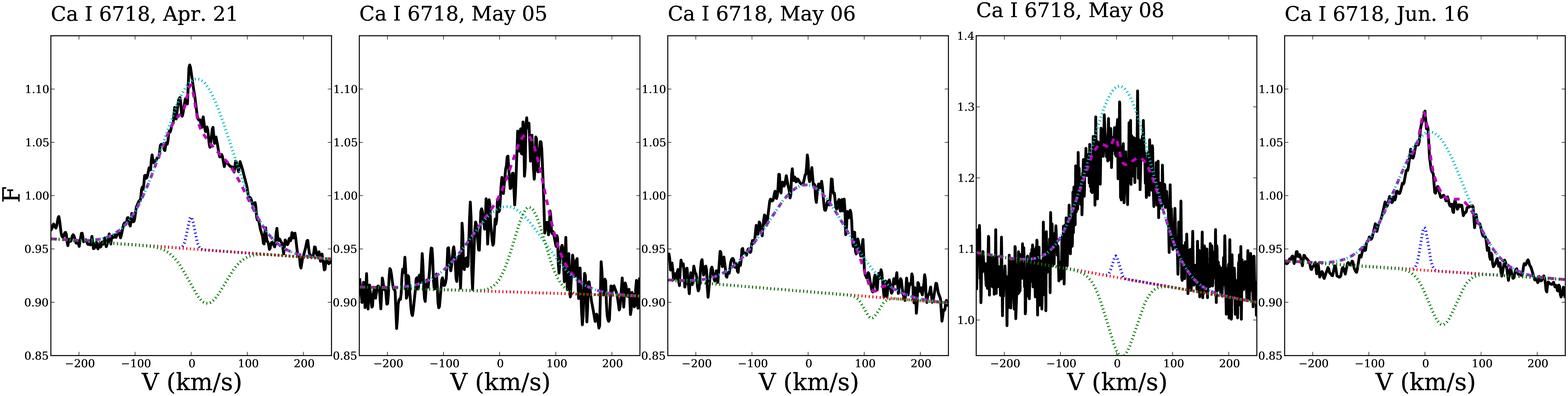}
\caption{Ca~I 6718\AA\ line: Results of the fit with three Gaussian components
for the different outburst epochs. The original data
is shown as the bold line, individual Gaussians
are marked by dotted lines, and the final fit is shown as a dashed line.   \label{CaI6718fit-fig}}
\end{figure*}

\begin{figure*}
   \centering
   \includegraphics[width=18cm]{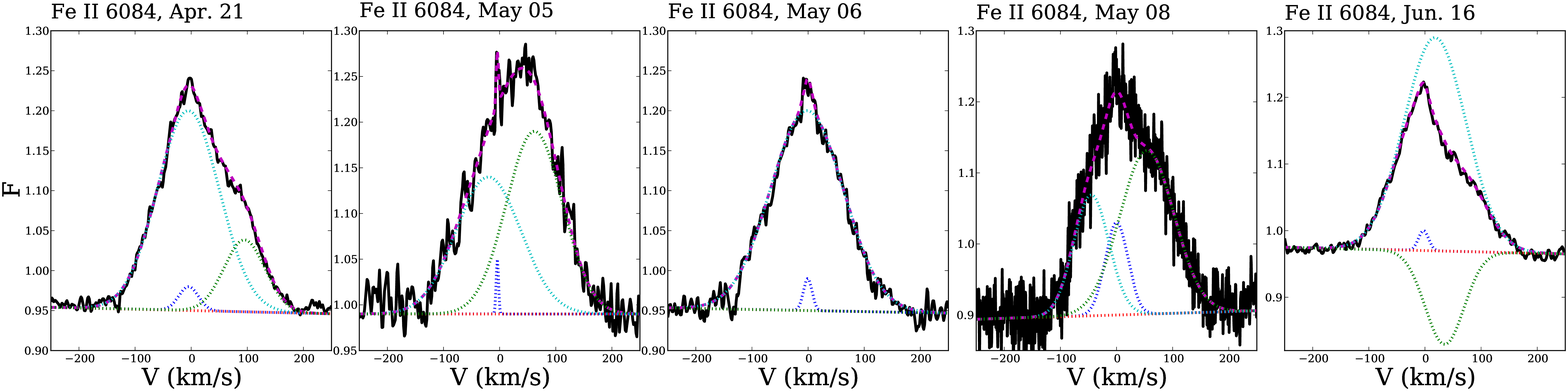}
\caption{Fe~II 6084\AA\ line: Results of the fit with three Gaussian components
for the different outburst epochs. The original data
is shown as the bold line, individual Gaussians
are marked by dotted lines, and the final fit is shown as a dashed line.   \label{FeII6084fit-fig}}
\end{figure*}

\begin{figure*}
   \centering
   \includegraphics[width=18cm]{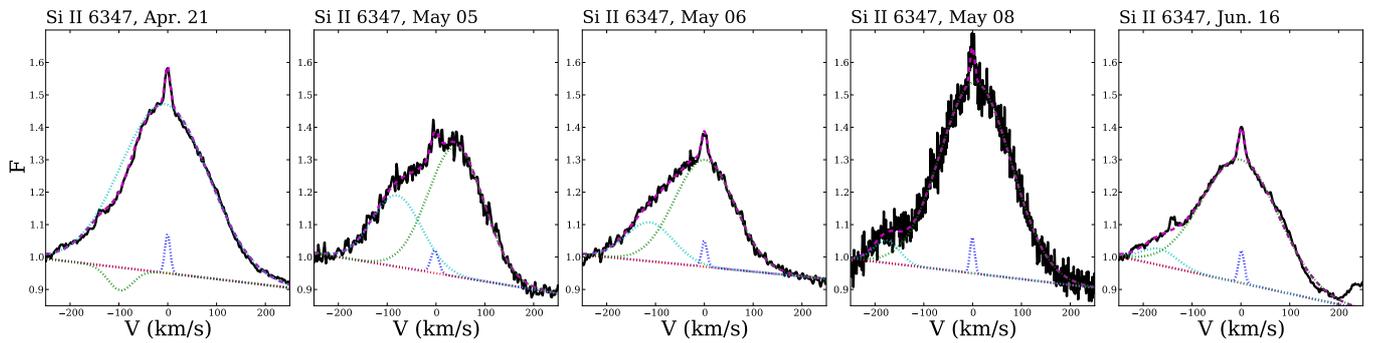}
\caption{Si II 6347\AA\ line: Results of the fit with three Gaussian components
for the different outburst epochs. The original data
is shown as the bold line, individual Gaussians
are marked by dotted lines, and the final fit is shown as a dashed line.  Note that the
blueshifted emission that appears on June 16 is probably due to contamination
by a weaker, blended line that became visible as the continuum and BC strength decreased. 
\label{SiII6347fit-fig}}
\end{figure*}

We describe here in detail the three Gaussian fits to the independent lines in the outburst, and
present a few more examples of the resulting best fits to some of the metallic lines
(see Figures \ref{MgI4702fit-fig} to \ref{SiII6347fit-fig}). The fits were done with
IRAF task $ngaussfit$, restricting the fitting range to $\pm$250 km/s (or less, in the case
of proximity to another feature). The fitting was done interactively, paying special attention
to the continuum and the presence of the NC. The model fitted to the data can be written 
as

\begin{equation}
	F_{total}(v) 
	= a + b v + 
	\sum_{i=1}^3 A_i e^{-\frac{(v-C_i)^2}{2 (\text{FWHM}_i/2 \sqrt{2 \ln2})^2}}, 
	\label{eq:3}
\end{equation}
where a and b are the continuum level and slope, A$_i$ are the amplitudes of the different Gaussian components, C$_i$ are
the Gaussian centers (in velocity), and FWHM$_i$ are the FWHM of the three Gaussian components.
The full collection of best fits to the lines unaffected by blending is shown in Table \ref{gaussfit-table}. 
While the BC of neutral and ionized metallic lines observed in May 2008 are
strongly consistent with two broad Gaussian components with variable velocities, 
the April and June observations in 2008 were not so uniquely defined. Fitting in terms of
a single broad component plus either a narrower blueshifted or redshifted absorption component
typically provides the best fits to the metallic lines (and would be a signature of wind and infall, 
respectively), but we cannot a priori rule out an interpretation in terms of 
two Gaussian emission components. In several cases during these two dates, a simple three Gaussian fit
to all the components in emission fails to reproduce the blueshifted and redshifted absorption
profiles observed, which reinforces the interpretation of the line profiles 
in terms of wind- and infall-related absorption. In addition, in some other cases, the fits are
not unique (e.g. Fe II line on April 21, see Table \ref{gaussfit-table}).

On the other hand, the H I and He I lines appear much more symmetric than
the metallic lines at all dates, and the asymmetries that could be potentially
interpreted as independent components do not show any clear dynamical pattern in the
different observations, which clearly distinguishes them from the metallic lines.

%\longtabL{A1}{
\begin{landscape}
\begin{longtable}{lccccccccccccccl}
\caption{\label{gaussfit-table} Gaussian fits. The different components are listed with the NC first, then ``bluer'', then ``redder". Only
lines with no evident blends are fitted. In the `Comments' column, we include a note if additional
components would be required to properly fit the line. `S/N' means that the line had relatively low S/N, so the
fit is uncertain and the Gaussian components relatively unconstrained. `C' means that the 
continuum was complicated (e.g., by the presence of nearby lines), so some extrapolation
was needed in order to do the fit. The rest of comments are self-explanatory.}\\
\hline\hline
Line &  Date & $\chi^2$ &rms &a & b  &A1 & C1& FWHM1  &A2 &C2& FWHM2  &A3& C3& FWHM3 & Comments\\ 
\hline
\endfirsthead
\caption{Continued.}\\
\hline\hline
Line &  Date & $\chi^2$ &rms &a & b  &A1 & C1& FWHM1  &A2 &C2& FWHM2  &A3& C3& FWHM3 & Comments\\ 
\hline
\endhead
\hline
{\bf IR H~I lines}\\
H~I~8598   & 04.21. & 9.6E-5 & 0.010 & 1.27 & -2.5E-5 & 0.04 & 6.9  & 43.   & 0.79 & 2.   & 193. & 0.07  & 107.  & 47. & \\
H~I~8750   & 04.21. & 0.001  & 0.032 & 1.45 & 1.8E-4  & 0.   & 0.   & 0.    & 1.29 & 1.   & 171. & 0.22  & 176.  & 201. & \\
H~I~8598   & 05.05. & 6.6E-4 & 0.026 & 0.85 & 8.4E-6  & -0.03& 11.1 & 40.  & 0.48 & 22. & 181.  & 0.09 & -110. & 187. & \\
H~I~8750   & 05.05. & 0.002  & 0.044 & 1.00 & 2.2E-5  & 0.002 & 10.0 & 24. & 0.34 & -23. & 170. & 0.50 & 47. & 204. & \\
H~I~8598   & 05.06. & 3.0E-4 & 0.017 & 1.15 & -1.5E-6 & -0.05 & -31. & 15. & 0.34 & -34.  & 215.  & 0.26  & 16. & 156. & \\ 
H~I~8750   & 05.06. & 0.001  & 0.034 & 1.40 & 3.5E-4  & 0.   & 0.    & 0.  & 0.88 & -4.   & 170.  & 0.    & 0.  & 0.   & \\
H~I~8598   & 06.16. & 7.9E-5 & 0.009 & 1.10 & -3.5E-5 & -0.08 & 43. & 43.  & 0.28  & -38. & 212.  & 0.40  & 24.  & 145. & \\
\hline
\end{longtable}
\end{landscape}
%}

\end{appendix}

\end{document}